\documentclass[aps,pre,preprint,groupedaddress,showpacs]{revtex4-1}
\usepackage[dvips]{graphicx}
\usepackage{textcomp}
\usepackage{amssymb}
\usepackage{dsdshorthand}

\newcommand{\mc}{\multicolumn}
\begin{document}

\title{
\Large\bf Restoring isotropy in a three-dimensional lattice model:
       The Ising universality class}

\author{Martin Hasenbusch}
\email[]{M.Hasenbusch@thphys.uni-heidelberg.de}
\affiliation{
Institut f\"ur Theoretische Physik, Universit\"at Heidelberg,
Philosophenweg 19, 69120 Heidelberg, Germany}

\date{\today}

\begin{abstract}
We study a generalized Blume-Capel model on the simple cubic lattice. 
In addition to the nearest neighbor coupling there is a next to next
to nearest neighbor coupling. In order
to quantify spatial anisotropy, we determine the correlation length
in the high temperature phase of the model for three different spatial
directions.
It turns out that the spatial anisotropy depends very little on the 
dilution, or crystal-field parameter $D$ of the model and is 
essentially determined by
the ratio of the nearest neighbor and the next to next to nearest
neighbor coupling. This ratio is tuned such that
the leading contribution to the spatial anisotropy is eliminated. 
Next we perform a finite size scaling (FSS) study to tune $D$ such that also
the leading correction to scaling is eliminated. Based on this FSS study,
we determine the critical exponents $\nu=0.62998(5)$ and $\eta=0.036284(40)$, 
which are in nice agreement with the more accurate results obtained by 
using the conformal bootstrap method. Furthermore we provide accurate results
for fixed point values of dimensionless quantities such as the Binder 
cumulant and for the critical couplings. These results provide the groundwork
for broader studies of universal properties of the three-dimensional Ising 
universality class.
\end{abstract}

\maketitle

\section{Introduction}
Studying spin models in the neighborhood of the critical temperature
numerically, the presence of corrections
to scaling hampers the extraction of universal quantities. The 
straight forward approach to reduce the effect of corrections to scaling
is to simulate larger and larger lattices. 
It is more economic to study a family of models, and tune one 
or more parameters of the family such that the amplitude of the leading 
correction vanishes. This idea dates back to \cite{ChFiNi,FiCh}, where it
is implemented by using high temperature series expansions.  
The idea had been picked up in finite size scaling (FSS) \cite{Barber}
studies using 
Monte Carlo simulations in refs. \cite{BlLuHe95,Ballesteros,KlausStefano}, 
where the universality class of the three-dimensional Ising model had been
studied.
The idea has been applied successfully to the XY \cite{Tibor,XY1,XY2,myClock}, 
the Heisenberg \cite{myO3O4,ourHeisen,myIco}, and the disordered Ising 
\cite{ourdilute}
universality classes in three dimensions, resulting in accurate estimates of
critical exponents. 
Note that the related improvement programme initiated by Symanzik
\cite{Sy83} 
is an indispensable building block in today's lattice QCD simulations. 
An open question is, whether this programme can be extended successfully to 
subleading corrections.  Here we do not answer this question in general but
consider one particular case.
We study a lattice model with a second order phase transition in the 
universality class of the three-dimensional Ising model.
We extend the idea of eliminating corrections to scaling to subleading
corrections that are caused by spatial anisotropy.

In the last years, the conformal bootstrap (CB) method brought enormous 
progress in the study of critical phenomena in three dimensions. In contrast
to previous methods, the starting point is not a Hamiltonian. Instead, 
conformal invariance and qualitative features of the fixed point are 
the basis of the analysis. The programme has provided highly accurate 
results for critical exponents and for operator product expansion coefficients.
For a recent review see, for example, ref. \cite{PoRyVi19}.
In particular, in the case of the three-dimensional Ising universality 
class, detailed information on correction exponents is provided. 
See table 2 of ref. \cite{Simmons-Duffin:2016wlq}.

In a finite size scaling study, the spatial anisotropy of the system leads to 
corrections that vanish like $L^{-\omega_{NR}}$, where 
$\omega_{NR} = 2.0208(12)$ \cite{ROT98,pisaseries} and $L$ is the linear 
size of the system. The scaling field method \cite{NewmanRiedel} predicts
a subleading correction with the correction  exponent
$\omega'=1.67(11) < \omega_{NR}$ for the three-dimensional Ising 
universality class. Based on this result, it seemed 
of little use in the numerical study to eliminate the spatial anisotropy by tuning the
parameters of the reduced Hamiltonian.  However the CB method,
consistent with the functional renormalization group (FRG), see for example
ref. \cite{LiVe04},  indicates that $\omega'=1.67(11)$ is an artifact 
of the scaling field method.
For a more detailed discussion see section \ref{Sec_corrections} below. 

Based on this observation it seems promising to study reduced Hamiltonians,
where in addition to the leading correction to scaling, the spatial 
anisotropy is eliminated to leading order. To this end we study the
Blume-Capel model on the simple cubic lattice, where
in addition to the nearest neighbor coupling, there is a third nearest 
neighbor coupling. This model has two parameters that can be tuned 
to remove corrections to scaling: The ratio of the two coupling constants and
the parameter $D$ that controls the density of vacancies. The definition 
of the model is given below in section \ref{themodel}.

The leading correction to scaling is eliminated by using a finite 
size scaling study similar to our previous work, see \cite{myIco} and 
references therein. In order to quantify the spatial anisotropy, we study the 
correlation length in the high temperature phase in different spatial 
directions.
Finite size scaling is less practical, since the rotational invariance
is not only broken at the microscopic scale by the lattice but also at
large length scales by the torus geometry of the lattice with periodic
boundary conditions. This can be seen in two-point correlation functions 
even at rather small distances, see for example \cite{Mythreepoint}.
Instead, we study the correlation length 
in the high temperature phase, where the parameters of the reduced Hamiltonian
are chosen such that $L \gg \xi$.  This way, the correlation 
functions at scales $\sim \xi$ are very little affected 
by the global torus geometry. 
In the high temperature phase of the Ising model and related
models, the correlation length can be determined very accurately 
by using a variance reduced estimator of the two-point correlation function 
that is associated with the cluster algorithm \cite{SwWa87,Wolff}. 

Based on the FSS analysis, we get very accurate estimates of the critical
exponents $\nu$ and $\eta$ that are fully consistent with the CB estimates.
Furthermore we get very accurate results for the inverse critical temperature, 
which is valuable input for future studies of the model discussed here.
Reduced spatial anisotropy should be, for example, helpful in the study of
interfaces
in the low temperature phase or the thermodynamic Casimir effect with 
non-trivial geometries.

Here we mostly delve into specifics of critical phenomena. 
For general reviews on critical phenomena and the renormalization 
group (RG) theory see, for example,  
\cite{WiKo,Fisher74,Fisher98,PeVi02}. 

The outline of the paper is the following: In section \ref{themodel} we 
define the model. In section \ref{Sec_corrections} follows a more detailed 
discussion on corrections
to scaling. In section \ref{Restoring} we determine the ratio of 
nearest and next to next to nearest neighbor couplings that restores 
isotropy to leading order. To this end we study the correlation length 
in different directions in the high temperature phase of the model. 
In section \ref{FSSsec}, by using FSS, we determine the value
$D^*$ of the dilution parameter, where leading corrections to scaling are
eliminated. 
Based on this FSS study we obtain accurate estimates of critical exponents.
Finally we summarize and conclude.

\section{The model}
\label{themodel}
We study a generalized Blume-Capel model on the simple cubic lattice, 
where in addition 
to the nearest neighbor coupling, there is a non-vanishing third nearest 
neighbor coupling. This model has been discussed for example in ref. 
\cite{BlLuHe95}. See in particular eq.~(2) of ref. \cite{BlLuHe95}.
For a vanishing external field, it is defined by the reduced Hamiltonian
\begin{equation}
\label{BlumeCapel}
H = -K_1 \sum_{<xy>}  s_x s_y - K_3 \sum_{[xy]}  s_x s_y 
  + D \sum_x s_x^2   \;\; ,
\end{equation}
where the spin $s_x$ might assume the values $s_x \in \{-1, 0, 1 \}$.
$x=(x^{(0)},x^{(1)},x^{(2)})$ denotes a site on the simple cubic lattice,
where $x^{(i)} \in \{0,1,...,L_i-1\}$. Furthermore,
$<xy>$ denotes a pair of nearest
and $[xy]$ a pair of next to next to nearest, or third nearest  neighbors on 
the lattice. In this study we consider $L_0=L_1=L_2=L$ and periodic 
boundary conditions throughout. Here we refer to $D$ as dilution parameter.
In the literature, $D$ is also denoted as crystal-field parameter.
The partition function is given by $Z = \sum_{\{s\}} \exp(- H)$, where 
the sum runs over all spin configurations. In the following we denote
the ratio of coupling constants as
\begin{equation}
q_3 = K_3/K_1  \;\;.
\end{equation}
For $q_3=0$, the model has been thoroughly studied in the literature. 
See \cite{MHcritical} and references therein.
In the limit $D \rightarrow - \infty$, the vacancies $s_x =0$ are 
completely suppressed, and the Ising model is recovered. For $D < D_{tri}$, 
the model undergoes a second order phase transition in the universality
class of the three-dimensional Ising model.  For $D > D_{tri}$, there
is a first order phase transition. Along the line of second order transitions,
the amplitude of leading corrections depends on the parameter $D$. It 
has been demonstrated numerically that there is a value $D^*$ of the parameter,
where leading corrections to scaling vanish.  In ref. \cite{MHcritical}
we find $D^* = 0.656(20)$, which 
is clearly smaller than $D_{tri} = 2.0313(4)$ \cite{DeBl04}.
For a more detailed discussion see ref. \cite{MHcritical}.

For the model, eq.~(\ref{BlumeCapel}), for $q_3 \ge 0$, we expect that there is
a critical plane given by $K_{1,c}(D,q_3)$ that is bounded by a line of 
tricritical transitions $D_{tri}(q_3)$. On the critical plane, there should
be a line $D^*(q_3)$, where leading corrections to scaling vanish.  There
should be also a line, where the isotropy is restored to leading order. 
It is best represented by $q_3^{iso}(D)$, since we expect that $q_3^{iso}(D)$ 
depends 
only little on $D$, which is confirmed by our numerical results discussed
below. These two lines might have a crossing, where both corrections to 
scaling vanish.  

In ref. \cite{BlLuHe95} as well as in the more recent 
papers \cite{BlShTa99,DengBloete03},  the Ising model, corresponding
to $D \rightarrow -\infty$, with nearest and 
next to next to nearest neighbor couplings had been studied.
It turns out that the amplitude of leading corrections to scaling depends on 
the ratio $q_3$. In particular, there is a value $q_3^*$, where leading 
corrections to scaling vanish. The authors of refs. 
\cite{BlLuHe95,BlShTa99,DengBloete03} performed a finite size scaling 
analysis based on the quantity $Q=<m^2>^2/<m^4>$, where $m$ is the 
magnetization. Note that $Q$ is the inverse of the Binder cumulant 
defined here, eq.~(\ref{Binder2j}) for $j=2$.
In table III of ref. \cite{DengBloete03} the estimates 
$b_1=0.097(2)$, $0.051(2)$, $0.0118(20)$, $-0.0180(20)$, and $-0.0480(20)$
for $q_3=0$, $0.1$, $0.2$, $0.3$, and $0.4$, respectively, are given,
where $b_1$ denotes the amplitude of the leading correction.
Interpolating linearly, we arrive at $q_3^*=0.24(1)$. 

Assuming that $D^*(q_3)$ is monotonically decreasing with increasing $q_3$,
the crossing of $q_3^{iso}(D)$ and  $D^*(q_3)$ exists if 
$q_3^{iso}(-\infty) \le q_3^*$.  Hence, as a first step of our numerical study, 
we determine $q_3^{iso}(-\infty)$.  

In the following we approach the critical line keeping $q_3$ constant.
Therefore we use the parameterization 
\begin{eqnarray}
\label{paramet}
K_1 &=& K  \;, \nonumber \\
K_3 &=& q_3 K \;.
\end{eqnarray}

\section{Corrections to scaling}
\label{Sec_corrections}
Field theoretic methods and high temperature 
series expansions and Monte Carlo simulations of lattice models 
give consistently for the leading correction to scaling 
exponent $\omega \approx 0.8$ for the three-dimensional Ising 
universality class. For a summary of results see, for example, 
table 19 of \cite{PeVi02}. 
The most accurate result $\omega=0.82968(23)$, is 
obtained by using the CB method \cite{Simmons-Duffin:2016wlq}.  Note that
in table 2 of Ref. \cite{Simmons-Duffin:2016wlq} dimensions $\Delta$ of 
operators are given. In the case of the leading correction, 
$\omega=\Delta_{\epsilon'}-3$ holds.

Before the advent of the CB method, information on subleading corrections
had been scarce. The $\epsilon$-expansion and perturbation theory in 
three dimensions fixed do not provide information on subleading corrections. 
In principle,
Monte Carlo renormalization group (MCRG) methods, see for example 
refs. \cite{Ma76,Sw76,Pa84,Ba92,Bl96}, are capable of producing such
results. However these are not given in the literature. 
Note that in these studies the error of the correction exponent
$\omega$ is considerably larger than that of the critical exponents.
Obtaining results for subleading corrections should be even harder.

In previous work, for example ref. \cite{MHcritical}, we assumed that the 
results obtained in ref. \cite{NewmanRiedel}
by using the scaling field method on subleading corrections to scaling 
are correct. It predicts a subleading correction with $\omega' = 1.67(11)$.  
This result
is in contradiction with results obtained by using  functional renormalization 
group methods. Depending on the approximation scheme that is used, results
$2.838 \le \omega' \le 3.6845$ are, for example, obtained in 
ref. \cite{LiVe04}. Recent work 
\cite{Simmons-Duffin:2016wlq}, using the conformal bootstrap method gives
$\omega'=\Delta_{\epsilon''}-3=3.8956(43)$.   It seems that 
$\omega' = 1.67(11)$ is an artifact of the scaling field method. 

There is a correction due to the fact that the simple cubic lattice 
breaks the spatial isotropy. This phenomenon can already be observed in the 
context of partial differential equations. See for example ref. \cite{PaKa05},
where the Laplacian on a square and a simple cubic lattice is discussed.  
These results directly apply to free field
theory on the lattice. Hence, for free field theory on the simple cubic lattice 
we get $\omega_{NR,free}=2$ and $q_{3,free}^{iso} = 1/8$. 

In the case of the three-dimensional Ising universality class one
gets $\omega_{NR} = 2.0208(12)$, see table 1 of \cite{pisaseries}, or 
 by using the CB method $\omega_{NR} = 
\Delta_{C_{\mu \nu \rho \sigma}} -3 =  2.022665(28)$, given in table 2 of 
Ref. \cite{Simmons-Duffin:2016wlq}.
Note that the value of the correction exponent differs only by little 
from the free field value. Also the value of $q_{3}^{iso}$ that 
we find below differs only by little from the free
field value. This fact is a bit surprising, since $q_{3}^{iso}$ should 
depend on the details of the model.

Our numerical analysis relies on the fact that amplitudes of 
corrections to scaling
are smooth functions of the parameters of the reduced Hamiltonian, 
as it is predicted by RG-theory. 
Furthermore, following  RG-theory, ratios of correction amplitudes
in different quantities, for the same type of correction, are universal
(For a discussion, see for example section 1.5 of Ref. \cite{PeVi02}). 
This fact in particular implies that corrections in different quantities
vanish at the same values, in our case 
$(q_3^{iso},D^*)$, of parameters of the reduced Hamiltonian.  

\section{Restoring isotropy}
\label{Restoring}
Our numerical study consists of two essentially separate parts. 
Following the hypothesis that $q_3^{iso}(D)$ depends only little on $D$, 
we first determine $q_3^{iso}$ for the Ising limit $D \rightarrow -\infty$.
Then we perform a preliminary finite size scaling study to get an 
estimate of $D^*(q_3^{iso}(-\infty))$.  For this estimate we  determine
again $q_3^{iso}$.  Since this estimate indeed differs very little from 
$q_3^{iso}(-\infty)$, we regard it as our final estimate.
In the second part of our study we perform an extensive FSS study 
to determine $D^*(q_3^{iso})$ accurately. 

The simulations in the high temperature phase of the Ising model were performed 
by using the single cluster algorithm \cite{Wolff}. In the case of the 
Blume-Capel model with finite $D$, local updates that allow the transition
from $s_x=0$ to $s_x=\pm 1$ and vice versa were used in addition. For a 
more detailed discussion of such a hybrid update scheme see for example 
section 5 of ref. \cite{MyVar}.  

\subsection{The correlation length in the high temperature phase}
In order to quantify spatial anisotropy, we determine the correlation length 
in three different directions of the lattice. Below we discuss how the 
correlation length is determined. We start with the definition of the 
basic quantities.

We define slice averages
\begin{equation}
 S(x_0) = \sum_{x_1,x_2} s_x \;,
\end{equation}
where the slice is perpendicular to the $(1,0,0)$-axis.  In addition, 
we consider slices perpendicular to the $(1,1,0)$ and the $(1,1,1)$-axis.
The corresponding slice averages are given by
\begin{equation}
\tilde S_{x_0}  = \sum_{x_1,x_2}   s_{x_0-x_1,x_1,x_2}
\end{equation}
and
\begin{equation}
\bar S_{x_0}  = \sum_{x_1,x_2}   s_{x_0-x_1-x_2,x_1,x_2} \;\;.
\end{equation}
Note that the arithmetics of the coordinates is understood modulo the linear 
lattice size $L$. The distance between adjacent slices is $d_s=1$, $2^{-1/2}$, 
and $3^{-1/2}$ for slices perpendicular to the $(1,0,0)$-, 
$(1,1,0)$- and the $(1,1,1)$-axis, respectively.

The slice correlation function is defined as 
\begin{equation}
 G(t) = \langle  S(x_0) S(x_0+t) \rangle \;\;. 
\end{equation}
Also here $x_0+t$ is understood modulo the linear lattice size $L$.
The correlation functions $\tilde G(t)$ and $\bar G(t)$ are defined 
analogously.

In our simulations, in order to reduce the statistical error,
we average over all $x_0$ and all directions equivalent to those given by the
$(1,0,0)$-, $(1,1,0)$- and the $(1,1,1)$-axis, respectively. The correlation 
function is determined by using the variance reduced estimator associated with 
the cluster algorithm \cite{SwWa87,Wolff}.

We define the effective correlation length
\begin{equation}
 \xi_{eff}(t) =  \frac{d_s}{\ln(G(t)/G(t+1))} \; , 
\end{equation}
where $L \gg t$ is assumed and $d_s$ is the distance between adjacent slices. 
To relax $L \gg t$ to some extent, we take 
the periodicity of the lattice into account. To this end we solve
numerically
\begin{eqnarray}
 G(t) &=& c \left (\exp\left(-\frac{d_s t}{\xi_{eff}(t)} \right)  +\exp\left(-\frac{d_s (L-t)}{\xi_{eff}(t)} \right)    \right) \;, \\
 G(t+1)&=& c \left (\exp\left(-\frac{d_s (t+1)}{\xi_{eff}(t)} \right)  +\exp\left(-\frac{d_s (L-t-1)}{\xi_{eff}(t)} \right)\right) 
\end{eqnarray}
with respect to $\xi_{eff}(t)$. For the Ising universality class in three dimensions,
in the high temperature phase, $\xi_{eff}(t)$ converges quickly as 
$t \rightarrow \infty$. See ref. \cite{MyVar} and references therein.

In a set of preliminary simulations, we determined the lattice size $L$ and 
distance $t$ that is 
needed to keep deviations from the desired limit $L \rightarrow \infty$ 
followed by $t \rightarrow
\infty$ at a size smaller than the statistical error.  
We conclude that $d_s t \simeq 2 \xi$ and 
$L \simeq 20 \xi$  is sufficient. In the following we take $\xi_{eff}(t)$ 
at $d_s t \simeq 2 \xi$ as estimate of the correlation length $\xi$.  
The direction is indicated by a subscript.

In order to quantify the spatial anisotropy, we study the ratios
\begin{equation}
\label{xiratios}
r_2=\frac{\xi_{(1,0,0)}}{\xi_{(1,1,0)}} \;\;,\; 
r_3=\frac{\xi_{(1,0,0)}}{\xi_{(1,1,1)}}
\end{equation}
in the neighborhood of the critical point.

\subsection{Numerical results for the Ising model and the Blume-Capel model
with nearest neighbor coupling only}
First  we simulated the standard Ising model in the high temperature phase.
The behavior of the correlation length is given by
\begin{equation}
 \xi = a (K_{c}-K)^{-\nu} \times 
       (1 + c (K_{c}-K)^{\theta} + d (K_{c}-K) + ...) \;,
\end{equation}
where $K_c$, $a$, $c$, and $d$ are non-universal constants. The critical
exponent of the correlation length is $\nu=1/y_t$, where $y_t$ is the thermal
renormalization group exponent. The 
correction exponent is $\theta=\nu \omega$. 
For numerical results of the second moment correlation length in the high 
temperature phase of the Ising model with $q_3=0$, see for example Appendix A
of ref. \cite{MyThermodynamic}. In ref. \cite{Landau18} the accurate 
estimate $K_c = 0.221654626(5)$ is given.

In the present study, we focus on $\xi < 10$. 
Our numerical results for the correlation length of the Ising model, $q_3=0$,
are summarized in table \ref{Isingxi}.  Note that in the case of the ratios
$r_2$ and $r_3$, the statistical correlation between the correlation lengths
in the different directions are properly taken into account by performing 
a Jackknife analysis.

\begin{table}
\caption{\sl \label{Isingxi}
Results for the correlation length of the Ising model with $q_3=0$.
In the first column we give the coupling $K$, in the second column we give
the linear lattice size $L$, and in the third column the correlation length
$\xi$ parallel to the $(1,0,0)$-axis. Then follow the ratios $r_2$ and $r_3$
defined in eq.~(\ref{xiratios}). 
}
\begin{center}
\begin{tabular}{lrlll}
\hline
\mc{1}{c}{$K$} &  \mc{1}{c}{$L$}  &  
\mc{1}{c}{$\xi_{(1,0,0)}$}  &
 \mc{1}{c}{$r_2$} & \mc{1}{c}{$r_3$} \\
\hline
 0.2  &  40 & 2.04147(4)  & 1.004922(7)  & 1.006606(8) \\
0.20944& 60 & 2.99993(4)  & 1.002281(4) &  1.003056(5)  \\
0.21376& 80 & 3.99868(4)  & 1.001281(3)  & 1.001715(3) \\
0.2161& 100 & 5.02713(13) & 1.000802(8)  & 1.001083(10) \\
0.21743&120 & 6.00095(9)  & 1.000562(4)  & 1.000757(5) \\
0.21896& 160& 8.01343(17) & 1.000321(6) & 1.000426(7) \\
\hline
\end{tabular}
\end{center}
\end{table}

We fitted the data with the Ansatz 
\begin{equation}
\label{vioansatz1} 
 r_i -1 = a \xi^{-x} \;,
\end{equation}
where $a$ and the exponent $x$ are free parameters. 
We refer to $\xi_{(1,0,0)}$ as $\xi$ to keep the notation simple.
The statistical error of $\xi$ is ignored for simplicity.
Fitting all data 
for $r_3$ we get $x=2.006(3)$ and $\chi^2$/d.o.f. $=0.26$.  Adding a 
correction term $\propto L^{-2}$ we get $x=2.016(12)$ and 
$\chi^2$/d.o.f. $=0.10$  instead.
We conclude that the exponent $x$ is consistent with the results for
$\omega_{NR}$ 
of refs. \cite{pisaseries,Simmons-Duffin:2016wlq}. However, our accuracy 
is by far lower than that of ref. \cite{Simmons-Duffin:2016wlq}.

Next we have simulated the Blume-Capel model on the simple cubic 
lattice with $q_3=0$
at $D=0.655$ at 8 values of $K$ that correspond to
$\xi \approx 2$, $3$, $4$, $5$, $6$, $7$, $8$, and $9$.  Our numerical results 
are given in table \ref{Blumexi}.

\begin{figure}
\begin{center}
\includegraphics[width=14.5cm]{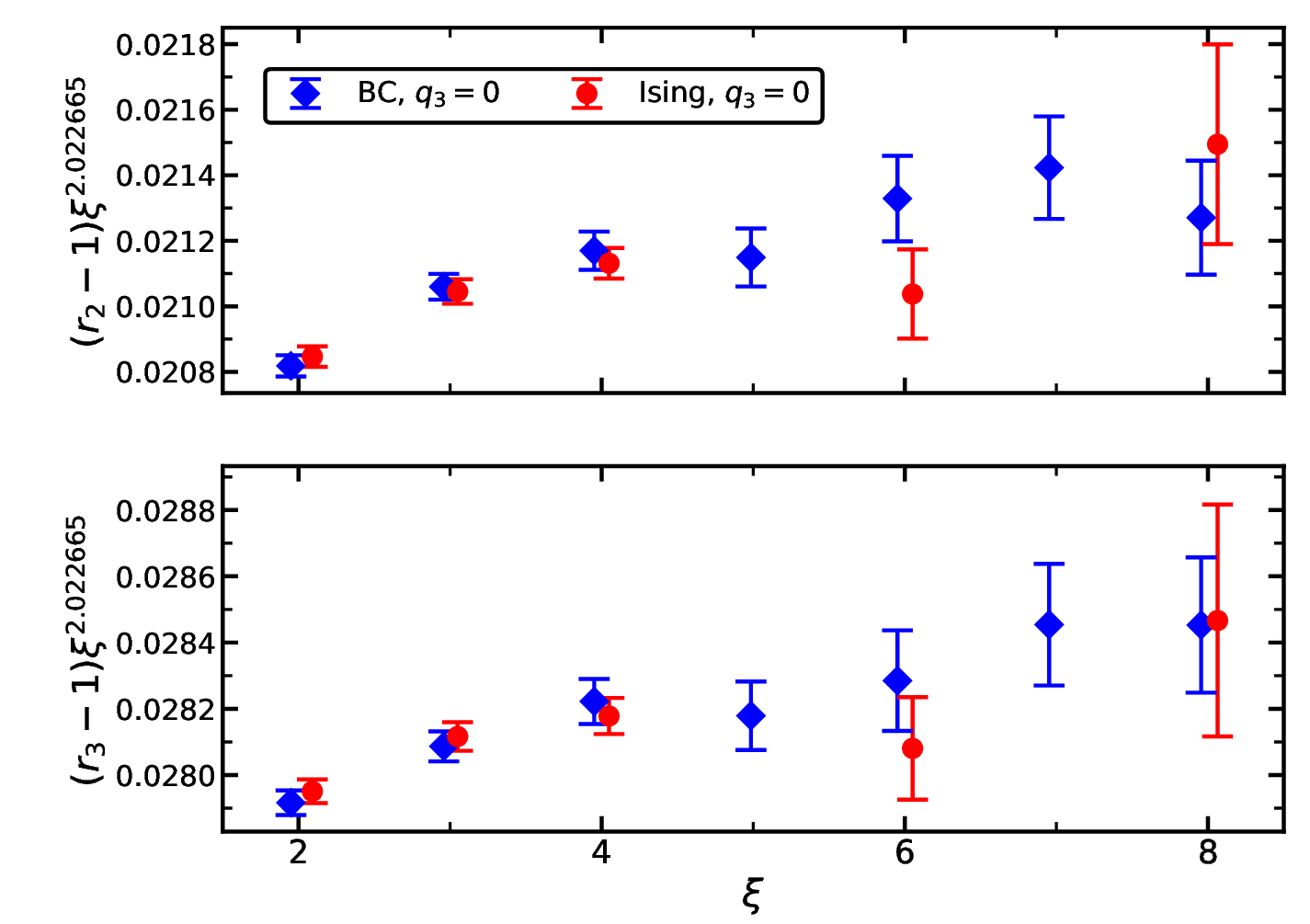}
\caption{\label{violF}
We plot $(r_2-1) \xi^{\omega_{NR}}$ (upper part) and 
$(r_3-1) \xi^{\omega_{NR}}$ (lower part) for the Ising model and the 
Blume-Capel model at $D=0.655$, both at $q_3=0$, versus the correlation length
$\xi$. 
Note that the values on the $x$-axis are slightly shifted to  reduce
the overlap of the Ising and the Blume-Capel data points. The two parts of 
the figure share the legend and the labeling of the $x$-axis. Note the 
different scales on the $y$-axis.
}
\end{center}
\end{figure}

\begin{table}
\caption{\sl \label{Blumexi}
We give results for the correlation length of the Blume-Capel model at 
$q_3=0$ and $D=0.655$.
In the first column we give the coupling $K$, in the second column we give
the linear lattice size $L$, and in the third column the correlation length
$\xi_{(1,0,0)}$ parallel to the $(1,0,0)$-axis. 
Then follow the ratios $r_2$ and $r_3$
defined in eq.~(\ref{xiratios}).
}
\begin{center}
\begin{tabular}{lrlll}
\hline
\mc{1}{c}{$K$} &  \mc{1}{c}{$L$}  & 
\mc{1}{c}{$\xi_{(1,0,0)}$}  &
 \mc{1}{c}{$r_2$} & \mc{1}{c}{$r_3$} \\
\hline
0.3568 &40 & 1.99990(5)    &  1.005124(8) &  1.006871(9) \\ 
0.3713 &60 & 3.00874(5)    &  1.002269(4) &  1.003026(5) \\ 
0.37721&80 & 4.00087(6)    &  1.001282(4) &  1.001709(4) \\ 
0.3804 &100 & 5.03495(8)    &  1.000804(3) &  1.001072(4) \\  
0.38217&120 & 6.00109(10)   &  1.000569(3) &  1.000754(4) \\ 
0.38337&140 & 7.00206(10)   &  1.000418(3) &  1.000555(4) \\  
0.3842 &160 & 8.00502(10)   &  1.000317(3) &  1.000424(3) \\ 
0.3848 &180 & 9.00819(13)   &  1.000251(3) &  1.000336(3) \\
\hline
\end{tabular}
\end{center}
\end{table}
Fitting all data for $r_3$ with $\xi \ge 3$ by using 
the Ansatz~(\ref{vioansatz1}) we get $x=2.010(4)$ and 
$\chi^2/$d.o.f. $=0.32$.
Fitting all data with an Ansatz containing a correction term $\propto L^{-2}$ 
we get $x=2.010(7)$ and $\chi^2/$d.o.f. $=0.32$. Fixing $x=2.022665$ in 
the Ansatz~(\ref{vioansatz1}),  we get very similar results for the
amplitude $a$ for both the Ising and the improved Blume-Capel model. 
We conclude, that the spatial anisotropy depends little on the 
amplitude of leading corrections to scaling, as we conjectured in
the beginning. This fact is illustrated in Fig. \ref{violF} were we
plot $(r_2-1) \xi^{\omega_{NR}}$ and $(r_3-1) \xi^{\omega_{NR}}$
versus the correlation length $\xi$.
The data for the two models fall essentially on top of each other.

\subsection{Determination of $q_3^{iso}$}
Next we determine $q_3^{iso}$ for the Ising model, corresponding
to $D \rightarrow - \infty$.  
Preliminary simulations give $q_3^{iso} \approx 2/15$.  
In order to get an 
accurate estimate of $q_3^{iso}$, we performed a number of simulations at 
$q_3=1/7, 2/15$,  and $1/8$ for a correlation length up to $\xi \approx 6$.  
Our results are summarized in table \ref{xxrI}.

\begin{table}
\caption{\sl \label{xxrI}
We give results for the correlation length $\xi_{(1,0,0)}$ 
and the ratios $r_2$ and $r_3$ of the Ising model with 
$q_3=1/7$, $2/15$, and $1/8$. 
$L$ is the linear lattice size and $K$ the coupling constant.
}
\begin{center}
\begin{tabular}{llrlll}
\hline
\mc{1}{c}{$q_3$} &
\mc{1}{c}{$K$} &  \mc{1}{c}{$L$}  & 
\mc{1}{c}{$\xi_{(1,0,0)}$}  &
 \mc{1}{c}{$r_2-1$} & \mc{1}{c}{$r_3-1$} \\
\hline
 $1/7$     & 0.1556 & 40 & 2.02430(7) & $-$0.000410(12) & $-$0.000695(14) \\
$1/7$  & 0.16938 &  100 & 5.02654(24) & $-$0.000043(14) & $-$0.000055(16) \\
\hline 
 $2/15$   & 0.158  & 40 &  2.03501(3)  & $-$0.000189(6) & $-$0.000372(7)  \\
$2/15$ & 0.1663 &  60 & 3.02940(4) & $-$0.000056(5) & $-$0.000074(5) \\
 $2/15$ & 0.16981 & 80 & 3.99996(6)  &  $-$0.000025(4) & $-$0.000028(5) \\
$2/15$ & 0.17175 &  100 & 5.00911(7) & $-$0.000008(4) & $-$0.000009(5) \\
$2/15$ & 0.172889 & 120 & 5.99953(9) & $-$0.000009(4) & $-$0.000004(5) \\
\hline 
 $1/8$ & 0.16 & 40 & 2.03256(3) & \phantom{$-$}0.000041(6) &$-$0.000050(7) \\
$1/8$ & 0.1686 & 60& 3.06724(4)  &\phantom{$-$}0.000043(5) & \phantom{$-$}0.000058(5)  \\
$1/8$ & 0.171951 & 80 & 3.99987(7) & \phantom{$-$}0.000025(5) & \phantom{$-$}0.000038(6) \\
$1/8$ & 0.1739 &100 & 5.00231(10)  & \phantom{$-$}0.000033(6)  & \phantom{$-$}0.000047(7) \\
 $1/8$  & 0.17506  & 120 & 5.99911(11)& \phantom{$-$}0.000009(5) &  \phantom{$-$}0.000020(6) \\
\hline
\end{tabular}
\end{center}
\end{table}

Furthermore, we  estimate $D^*$ for $q_3=2/15$. To this end, 
we performed a FSS study focussing on
$U_4$ at $Z_a/Z_p=0.5425$. For the definition of the Binder cumulant $U_4$ and 
the ratio of partition functions $Z_a/Z_p$ see section \ref{FSSsec} below.
 Here we simulated lattices up to the linear
size $L=32$. We used $U_{4,Z_a/Z_p=0.5425}^* \approx 1.60357$ obtained in
section VI of ref. \cite{MHcritical} as input.
We find $D^* \approx -0.43$.
Based on this preliminary result, we performed simulations at $D = -0.43$ for 
$q_3=2/15$ and $1/8$ in the high temperature phase. The value of $K$ is 
tuned such that the correlation length assumes the values $\xi \approx 2$, $3$,
$4$, $5$, $6$, $7$, and $8$.  Our results are summarized in table \ref{xxrB}.

\begin{table}
\caption{\sl \label{xxrB}
We give results for the correlation length $\xi_{(1,0,0)}$
and the ratios $r_2$ and $r_3$ of the Blume-Capel model at $D=-0.43$ with
$q_3=2/15$, and $1/8$. $L$ is the linear lattice size and $K$ the  
coupling.
}
\begin{center}
\begin{tabular}{llrlll}
\hline
\mc{1}{c}{$q_3$} &
\mc{1}{c}{$K$} &  \mc{1}{c}{$L$}  & 
\mc{1}{c}{$\xi_{(1,0,0)}$}  &
 \mc{1}{c}{$r_2-1$} & \mc{1}{c}{$r_3-1$} \\
\hline
2/15 & 0.2037 & 40 & 1.98782(3) &  $-$0.000222(7)   & $-$0.000431(7) \\
2/15 & 0.2144 & 60 & 3.00063(3) &  $-$0.0000699(27) & $-$0.0001043(31) \\ 
2/15 & 0.2188 & 80 & 4.00152(3) &  $-$0.0000344(19) & $-$0.0000494(22) \\ 
2/15 & 0.2211 &100 & 5.00263(4) &  $-$0.0000208(19) & $-$0.0000288(22) \\  
2/15 & 0.22247&120 & 6.00387(5) &  $-$0.0000140(19) & $-$0.0000227(22) \\ 
2/15 & 0.223356&140& 6.99959(5) &  $-$0.0000116(18) & $-$0.0000166(22) \\ 
2/15 & 0.223971&160& 7.99879(6) &  $-$0.0000067(18) & $-$0.0000081(21) \\
\hline
1/8 & 0.2065 &  40 & 2.00215(3) & \phantom{$-$}0.0000136(52) &$-$0.0001107(60) \\ 
1/8 & 0.217  &  60 & 2.99516(3) & \phantom{$-$}0.0000307(34) &\phantom{$-$}0.0000290(39) \\
1/8 & 0.22147&  80 & 4.00089(4) & \phantom{$-$}0.0000275(28) & \phantom{$-$}0.0000290(33) \\ 
1/8 & 0.2238 & 100 & 5.00716(4) & \phantom{$-$}0.0000168(18) & \phantom{$-$}0.0000215(22) \\
1/8 & 0.22517& 120 & 5.99963(5) & \phantom{$-$}0.0000116(19) & \phantom{$-$}0.0000120(22) \\ 
1/8 & 0.226065&140 & 6.99452(5) & \phantom{$-$}0.0000090(19) & \phantom{$-$}0.0000120(22) \\ 
1/8 & 0.226687&160 & 7.99376(6) & \phantom{$-$}0.0000070(19) & \phantom{$-$}0.0000078(22) \\ 
\hline
\end{tabular}
\end{center}
\end{table}

\begin{figure}
\begin{center}
\includegraphics[width=14.5cm]{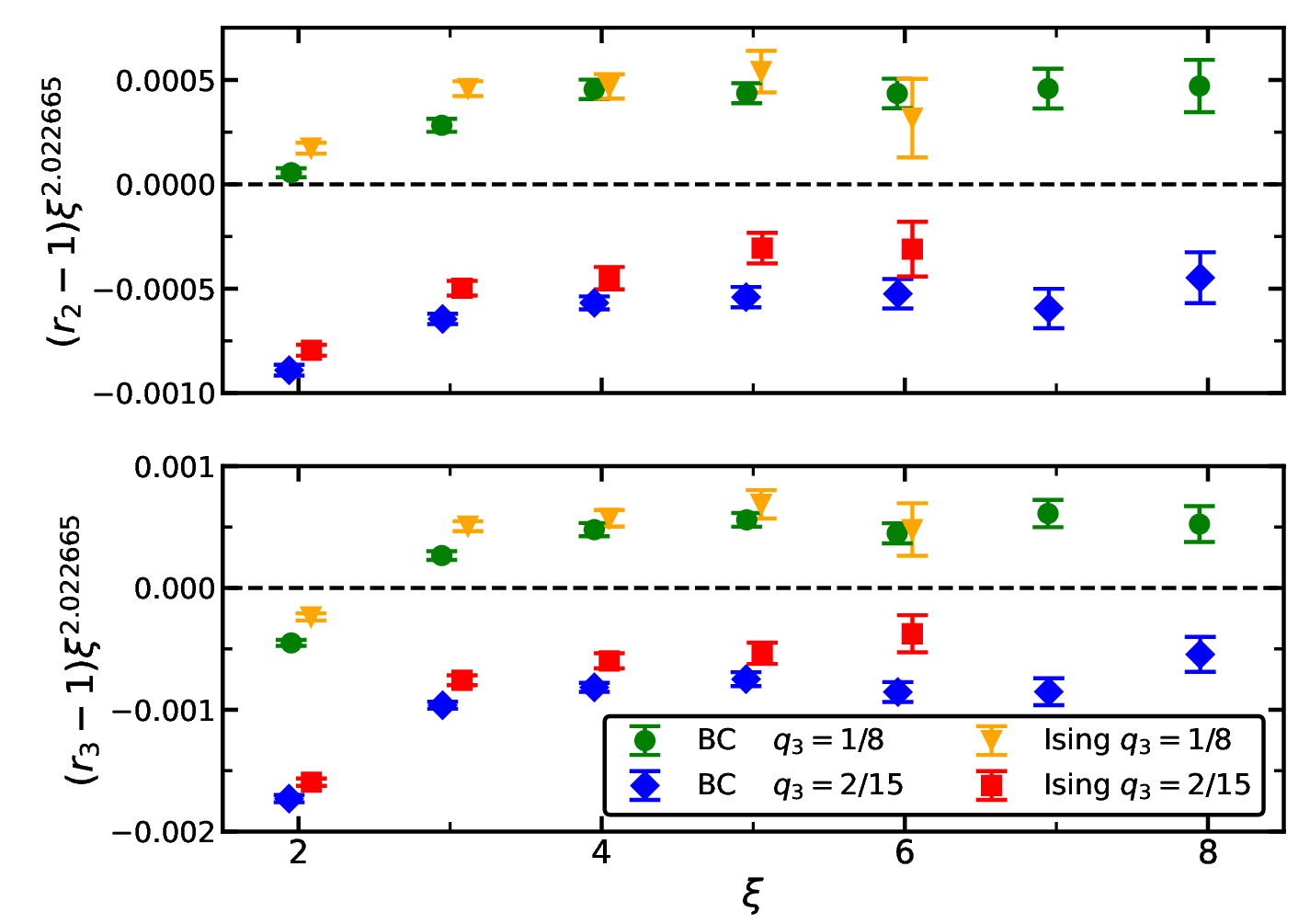}
\caption{\label{vanish3}
We plot $(r_2-1) \; \xi^{2.022665}$ (upper part) and  
$(r_3-1) \; \xi^{2.022665}$ (lower part) versus the correlation length 
$\xi$ for the Ising model and the  Blume-Capel model at $D=-0.43$ 
at $q_3=2/15$ and $1/8$.
Note that the values on the $x$-axis are slightly shifted to reduce
the overlap of the Ising and the Blume-Capel model data points.
The two parts of the figure share the legend and the labeling of the $x$-axis.
Note the different scales on the $y$-axis.
}
\end{center}
\end{figure}

In Fig. \ref{vanish3} we plot $(r_2-1) \xi^{\omega_{NR}}$ and 
$(r_3-1) \xi^{\omega_{NR}}$ versus the 
correlation length $\xi$. With increasing $\xi$
the values of $(r_2-1) \xi^{\omega_{NR}}$ and $(r_3-1) \xi^{\omega_{NR}}$
seem to approach a constant for both  models and both values of $q_3$ we 
simulated at. It seems obvious that $1/8 < q_3^{iso} < 2/15$ for both models.
The values of $(r_2-1) \xi^{\omega_{NR}}$ and
$(r_3-1) \xi^{\omega_{NR}}$ are slightly larger for the Ising 
model, suggesting that $q_3^{iso}$ is slightly larger for the Ising model
than for the Blume-Capel model at $D=-0.43$. 

In order to obtain a numerical estimate of $q_3^{iso}$ for the 
Blume-Capel model at $D=-0.43$ we performed fits with the
Ans\"atze
\begin{equation}
\label{devifit1}
 r_3-1 = a \xi^{-\omega_{NR}} 
\end{equation} 
and 
\begin{equation}
\label{devifit2}
 r_3-1 = a \xi^{-\omega_{NR}}  + b \xi^{-\omega_{NR}'} \;,
\end{equation} 
where we have fixed $\omega_{NR}=2.022665$. In the case of the
correction term we took either $\omega_{NR}'= 6.42065-3=3.42065$, see table 2  
of ref. \cite{Simmons-Duffin:2016wlq}, or the ad hoc choice $\omega_{NR}'= 4$.
For example, with the Ansatz~(\ref{devifit2}) and $\omega_{NR}'= 3.42065$, 
taking $\xi \gtrapprox 3$ we get $a=-0.00062(7)$ and $0.00073(7)$ for 
$q_3=2/15$ and $1/8$, respectively. Note that $q_3^{iso}$ is defined as
the zero of $a$. Linearly interpolating we get $q_3^{iso}=0.1295(3)$.
Based on the fits that we performed by using 
the Ans\"atze~(\ref{devifit1},\ref{devifit2})  we quote
\begin{equation}
\label{q3iso}
q_3^{iso} = 0.129(1) 
\end{equation}
as final result for the Blume-Capel model at $D=-0.43$.
It is chosen such that the estimates, including their respective error bars,
obtained by performing these fits are covered. We did not repeat this analysis
for $r_2$. However just comparing the upper and lower part of 
Fig. \ref{vanish3} by eye, it is clear that the outcome of such an analysis 
will be very similar.

Below we perform a thorough FSS study, resulting in $D^* = -0.380(5)$ for 
$q_3=0.129$. Since the difference of
$q_3^{iso}$ for the Blume-Capel model at $D=-0.43$ and the Ising model is small,
we regard the result, eq.~(\ref{q3iso}), as valid for the revised estimate 
of $D^*$ and abstain from simulating again in the high temperature phase 
of the Blume-Capel model at $D^* = -0.38$. 

From Fig. \ref{violF} we read off that $(r_3-1) \xi^{\omega_{NR}} \approx 0.029$
for the Ising model and the Blume-Capel model at $D=0.655$ both at $q_3=0$ 
in the limit $\xi \rightarrow \infty$.
Taking the results of the fits discussed above for the amplitude of $r_3-1$ at
$q_3=1/8$  and $2/15$  we get 
$\mbox{d}[(r_3-1) \xi^{\omega_{NR}}]/\mbox{d} q_3 \approx 0.0013/(1/8-2/15)
=-0.156$ at $q_3=q_3^{iso}$.
Hence the error given in eq.~(\ref{q3iso}) means that for $q_3 = 0.129$,
the leading violation of spatial isotropy is suppressed at least
by a factor of about $0.029/(|-0.156| \times 0.001) \approx 180$ 
compared with $q_3=0$.

To get an idea of the statistics of our simulations let us briefly discuss 
the runs for $L=160$, $q_3=2/15$, and $D=-0.43$.  
In total we performed  about $3.6 \times 10^7$ 
update cycles. Each cycle consists of one sweep with the local update 
algorithm followed by 12000 single cluster updates. The number of 
single cluster updates is chosen such that this number times the average
size of a cluster roughly equals half of the volume $L^3$ of the lattice.
For parallelization, we performed 400 separate runs. 
For each run, we performed 10000 update cycles
for equilibration.  In total these runs took about 6 month of CPU time on
a single core of an AMD EPYC$^{TM}$ 7351P CPU. 
In the simulations discussed in this section, we 
used the SIMD-oriented Fast Mersenne
Twister (SFMT) algorithm \cite{twister} as random number generator.

Throughout this work, least square fits were performed by using the function 
\verb+curve_fit()+
contained in the SciPy  library \cite{pythonSciPy}. 
Plots were generated by using the Matplotlib library \cite{plotting}.

\section{Finite size scaling study}
\label{FSSsec}
In the second part of our numerical study we accurately determine 
$D^*$ for $q_3=0.129$. The outline of the study follows closely our recent 
studies \cite{myClock,myIco}. Therefore we abstain from a detailed discussion
of the theoretical background.  Below we define the quantities that we 
measure during the simulation. It follows a brief discussion of the simulations
that we performed. First we analyze the dimensionless quantities to locate
$D^*$ and get accurate estimates of $K_c$ for several values of $D$ 
close to $D^*$.  Next we obtain accurate estimates of the critical 
exponents $\eta$ and $\nu$ by analyzing the behavior of the magnetic
susceptibility and the slopes of dimensionless quantities.

\subsection{The quantities studied in finite size scaling}
\label{FSSquantities}
The magnetic susceptibility $\chi$ for a vanishing magnetization
and the second moment correlation length $\xi_{2nd}$ are defined as
\begin{equation}
\label{chidef}
\chi  =  \frac{1}{V} \,
\biggl\langle \Big(\sum_x s_x \Big)^2 \biggr\rangle
\end{equation}
and
\begin{equation}
\xi_{2nd}  =  \sqrt{\frac{\chi/F-1}{4 \sin^2 \pi/L}} \;,
\label{xidef}
\end{equation}
where
\begin{equation}
F  =  \frac{1}{V} \, \biggl\langle
\Big|\sum_x \exp\left(i \frac{2 \pi x_1}{L} \right)
         s_x \Big|^2
\biggr\rangle
\end{equation}
is the Fourier transform of the correlation function at the lowest
non-zero momentum. The Binder cumulant $U_4$ and its generalizations
$U_{2j}$ are defined as
\begin{equation}
\label{Binder2j}
U_{2j} = \frac{\langle (m^2)^j\rangle}{\langle m^2 \rangle^j} \;,
\end{equation}
where $m = \frac{1}{V} \, \sum_x s_x$ is the magnetization of
the system. 
Furthermore, we study the ratio of
partition functions $Z_a/Z_p$, where $a$ denotes a system with anti-periodic
boundary conditions in one of the directions and periodic ones in the remaining
two directions, while $p$ denotes a system with  periodic boundary conditions
in all directions. This quantity is computed by using the cluster algorithm.
For a discussion see Appendix A 2 of ref. \cite{XY1}.

The second moment correlation length $\xi_{2nd}$, the Binder cumulant $U_4$, 
its generalizations and the ratio of partition functions $Z_a/Z_p$
are dimensionless quantities or phenomenological couplings. In the 
following we denote these quantities by $R_i$.  We obtain the critical 
exponent $\nu$ from the behavior of the slope of dimensionless quantities
\begin{equation}
 S_{R_i} = \frac{\partial S_{R_i}}{\partial K} \;.
\end{equation}
In the analysis
discussed below, we need the quantities as a function of $K$ in some 
neighborhood of the value $K_{sim} \approx K_c$ of $K$ that 
is used in the simulation. To this end, we compute the Taylor coefficients 
of the observables around $K_{sim}$ up to third order.

For a discussion of corrections that are caused by the observable itself
see for example section 4 of ref. \cite{SalasSokal20}.  The authors discuss
the two-dimensional Ising model on the square lattice with periodic 
boundary conditions. The arguments brought forward should also apply 
to the present case. In particular it is noted that one has to take 
into account the analytic background of the magnetic susceptibility.
This leads to a correction in $U_{2j}$ and $\xi_{2nd}/L$ proportional to
$L^{-(2-\eta)}$. In the case of $\xi_{2nd}/L$, there are in addition 
corrections that are proportional to $L^{-2}$.
The ratio of partition functions has only corrections that decay exponentially
in the linear lattice size. 

\subsection{The simulations}
The simulations are performed by using a hybrid of local updates, single
cluster updates \cite{Wolff} and the wall cluster update \cite{KlausStefano}.
For each measurement, we performed one sweep with the local update, $L/4$
single cluster updates, and one wall cluster update. 
For a more detailed discussion of similar hybrid update schemes see for example
refs. \cite{myClock,myIco}.
We simulated the model for $q_3=0.129$ at $D=-0.3$, $-0.35$, 
$-0.38$, $-0.4$, $-0.42$, and $-0.46$. We simulated at good approximations of 
$K_{c}(D,q_3)$. These estimates were successively improved,  while increasing 
the linear lattice size that is simulated.

For all values of $D$ that we consider, we simulated the linear lattice sizes 
$L=6$, $7$, $8$, ..., $15$, $16$, $18$, $20$, ...,
$32$, $36$, $40$, $48$, $56$, ..., $72$. For $D=-0.3$, we simulated $L=120$ in 
addition. 
For $D=-0.35$, $-0.4$, and $-0.42$ we simulated $L=80$, $100$, and $120$
in addition. In the case of $D=-0.38$,
we simulated $L=80$, $100$, $120$, and $200$ in addition. 

In total we have spent the equivalent of about 90 years of CPU time on a 
single core of an AMD EPYC$^{TM}$ 7351P CPU.
To give the reader an idea of the statistics of our simulations: In the case of 
$D=-0.38$ we performed about $6.7 \times 10^9$ measurements for $L=20$. 
This number decreases to $1.5 \times 10^8$  measurements for $L=200$.
As random number generator, we used either the 
SIMD-oriented Fast Mersenne Twister (SFMT) algorithm \cite{twister} or
a modified KISS generator. A few simulations 
for $D=-0.38$ have been performed by using L\"uscher's ranlux generator
\cite{ranlux} for comparison. For more details see Appendix \ref{append}.
Analysing data and in particular estimating errors of the final results
for critical exponents and other quantities of interest, we follow a cautious
approach that we adopted over the years. It is spelled out for 
example in section V. of Ref. \cite{myClock}. Essentially, we perform a
number of different fits that we consider as reasonable. Then the final
result and its error bar are chosen such that the results of these fits, 
including their respective error bars are covered. This obviously leads in
general to a larger error bar compared with selecting one preferred fit
and taking its result and error bar as the final one.

\subsection{Dimensionless quantities}
\label{dimless}
In a first step we performed a joint fit of the dimensionless quantities
$Z_a/Z_p$, $\xi_{2nd}/L$, $U_4$, and $U_6$ for all values of $D$ considered.

As Ansatz we use
\begin{equation}
\label{allRansatz1}
 R_i(K_c,L) = R_i^* + b_i(D) L^{-\omega} + c_i(D) L^{-\epsilon_1} + 
  d_i(D) L^{-\epsilon_2}  \;.
\end{equation}
We have omitted 
corrections $c b_i^2(D) L^{-2 \omega} $ and higher powers, since 
$b_i(D)$ is assumed to be small for the values of $D$ that we consider.
In this section we have fixed $\omega=0.82968$, 
ref. \cite{Simmons-Duffin:2016wlq}.

In the case of $\xi_{2nd}/L$, $U_4$, and $U_6$ we expect that there are 
corrections due to the analytic background of the magnetic susceptibility.
Hence $\epsilon_1= 2 - \eta$.  In the case of $\xi_{2nd}/L$ there is in 
addition $\epsilon_2 = 2$, as discussed in section \ref{FSSquantities}. 
We assume that corrections due to the violation of the rotational invariance
can be ignored here. As a check, in the case of $Z_a/Z_p$, we assume one 
subleading correction with $\epsilon_1=2.022665$. 

The renormalization group predicts that the ratio $b_i(D)/b_j(D)$ does 
not depend on $D$. In our fits, we used different parameterizations of 
$b_i(D)$. For example, the linear approximation
\begin{equation}
\label{linearb}
 b_i(D) = a_i (D-D^*) \;,
\end{equation}
where $a_i$ and $D^*$ are free parameters. As check we added a quadratic 
term 
\begin{equation}
\label{quadraticb}
 b_i(D) = a_i [(D-D^*) + c (D-D^*)^2] \;.
\end{equation}
The coefficients of subleading corrections are assumed either to be
constant or linearly dependent on $D$.  In a preliminary stage of the 
analysis we performed a number of fits using different Ans\"atze of the 
type discussed above, including different subsets of values of $D$.
Note that by varying the range of $D$, we probe the validity of approximations
such as eqs.~(\ref{linearb},\ref{quadraticb}).
Motivating our final results, we focus on three different Ans\"atze that
we specify below.
Note that these three fits essentially cover the range of results that
we considered as reasonable in the preliminary stage of the analysis.

Fit 1: We include four values of $D$: $D=-0.35$, $-0.38$, $-0.4$, and $-0.42$. 
We parameterize the amplitude of leading corrections to scaling by using 
eq.~(\ref{quadraticb}). The coefficients of corrections related to the 
analytic background of the magnetic susceptibility are approximated by a 
linear function of $D$. All other coefficients of subleading corrections
are assumed to be constant.

Let us summarize the free parameters of the fit: 
$K_c$ for each value of $D$, $R^*_i$ for each dimensionless quantity, 
$D^*$, $a_i$, eq.~(\ref{quadraticb}), for each dimensionless quantity, 
$c$, eq.~(\ref{quadraticb}), two coefficients for each of $U_4$, $U_6$, and
$\xi_{2nd}/L$ for the correction related to the analytic background of the
magnetic susceptibility, one coefficients for the second subleading 
correction of $\xi_{2nd}/L$, and one coefficient for probing a possible 
correction $\propto L^{-2.022665}$ in $Z_a/Z_p$. 

Fit 2: We use the same Ansatz as for fit 1. In contrast to fit 1, 
we include all six values of $D$, where we simulated at. 

Fit 3:  We use the same data set as for fit 2. We use the same 
approximations for the coefficients in eq.~(\ref{allRansatz1}) as 
in fit 1 and 2. In contrast to fit 1 and 2, we add an additional 
correction term $e_i L^{-\epsilon_3}$, where now $\epsilon_3$ is a free 
parameter of the fit. It is assumed to be the same for all four dimensionless
quantities. In the Ansatz, $e_i$ does not depend on $D$.

In our fits, we include all data with a linear lattice size $L \ge L_{min}$.
Since corrections decrease with increasing $L$, the fits should become better, 
up to statistical fluctuations, with increasing $L_{min}$. In the following, 
we always plot results of the fits versus the minimal lattice size $L_{min}$. 
Let us discuss the results of the fits in detail: 

In the case of fit 1 we get $\chi^2/$d.o.f. $=4.68$, $1.73$, $1.22$, and 
$1.01$ for $L_{min} = 6$, $7$, $8$, and $9$, respectively. For $L_{min} \ge 10$
we get $\chi^2/$d.o.f. slightly smaller than one. The numbers for fit 2
look similar: We get $\chi^2/$d.o.f. $=4.54$, $1.68$, $1.22$, and 
$1.02$ for $L_{min} = 6$, $7$, $8$, and $9$, respectively. Again, 
for $L_{min} \ge 10$ we get $\chi^2/$d.o.f. slightly smaller than one.
In the case of fit 3, we get $\chi^2/$d.o.f.  $=1.04$  
for $L_{min} = 6$. For $L_{min} \ge 7$
we get $\chi^2/$d.o.f. slightly smaller than one.

In the case of fit 3 we get $\epsilon_3 \approx  5$, where the error bar 
is smaller than 1 only for $L_{min} \le 9$. We should be cautious in 
interpreting this result, since it is essentially based only on a few small
lattice sizes that discriminate fit 2 and fit 3. Certainly we can not 
exclude a correction with a smaller correction exponent and a small amplitude.

In the figures below we show data points for a $p$-value $p > 0.01$ only.
Corresponding to the $\chi^2$/d.o.f. discussed above, $p$ gets rapidly larger 
than this value, with increasing $L_{min}$.

\begin{figure}
\begin{center}
\includegraphics[width=14.5cm]{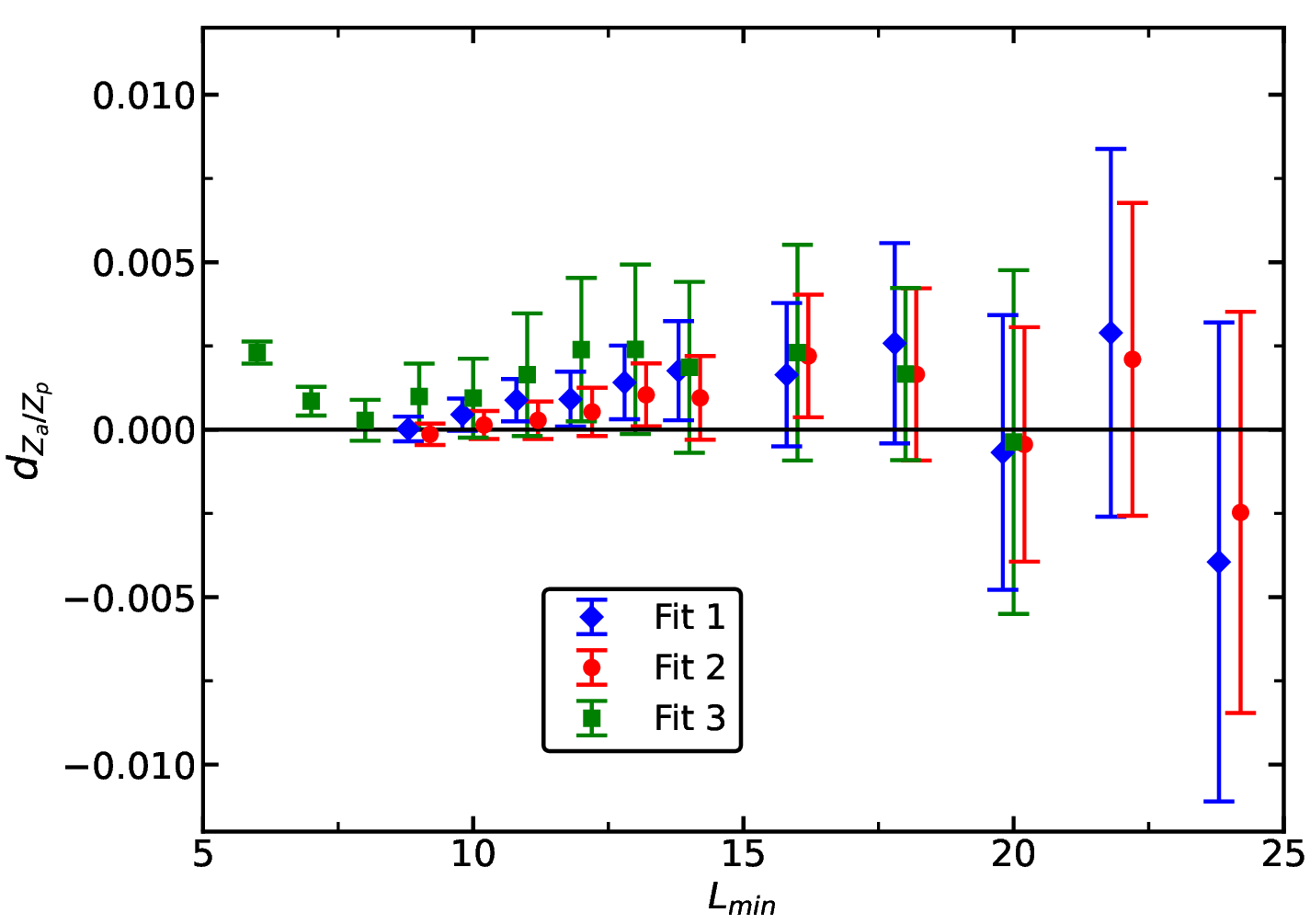}
\caption{\label{ZaZpcor}
Numerical estimates of the amplitude $d_{Z_a/Z_p}$ of corrections 
$\propto L^{-2.022665}$ 
in $Z_a/Z_p$ as a function of the minimal lattice size $L_{min}$. These 
estimates are obtained from the fits 1, 2, and 3, which
are discussed in the text. 
Note that the values on the $x$-axis are slightly shifted to reduce overlap
of the symbols.
}
\end{center}
\end{figure}

In Fig. \ref{ZaZpcor}  we give our numerical results for the amplitude
of the correction 
$\propto L^{-\omega_{NR}}$ of $Z_a/Z_p$. We find that it is
compatible with zero. For comparison, we have reanalyzed the  data of 
\cite{MHcritical} 
for the Blume-Capel model at $q_3=0$ for $D=0.641$, $0.655$, and $\ln 2$. 
We used the final estimates of the fixed point values of the dimensionless 
quantities obtained here as input, taking into account their covariances.
As estimate of the amplitude of the correction $\propto L^{-\omega_{NR}}$ of 
$Z_a/Z_p$ we find $d=-0.047(5)$. In the case of the 
other three quantities it is impossible to disentangle the 
correction $\propto L^{-\omega_{NR}}$ from the analytic background of the 
magnetization.

\begin{figure}
\begin{center}
\includegraphics[width=14.5cm]{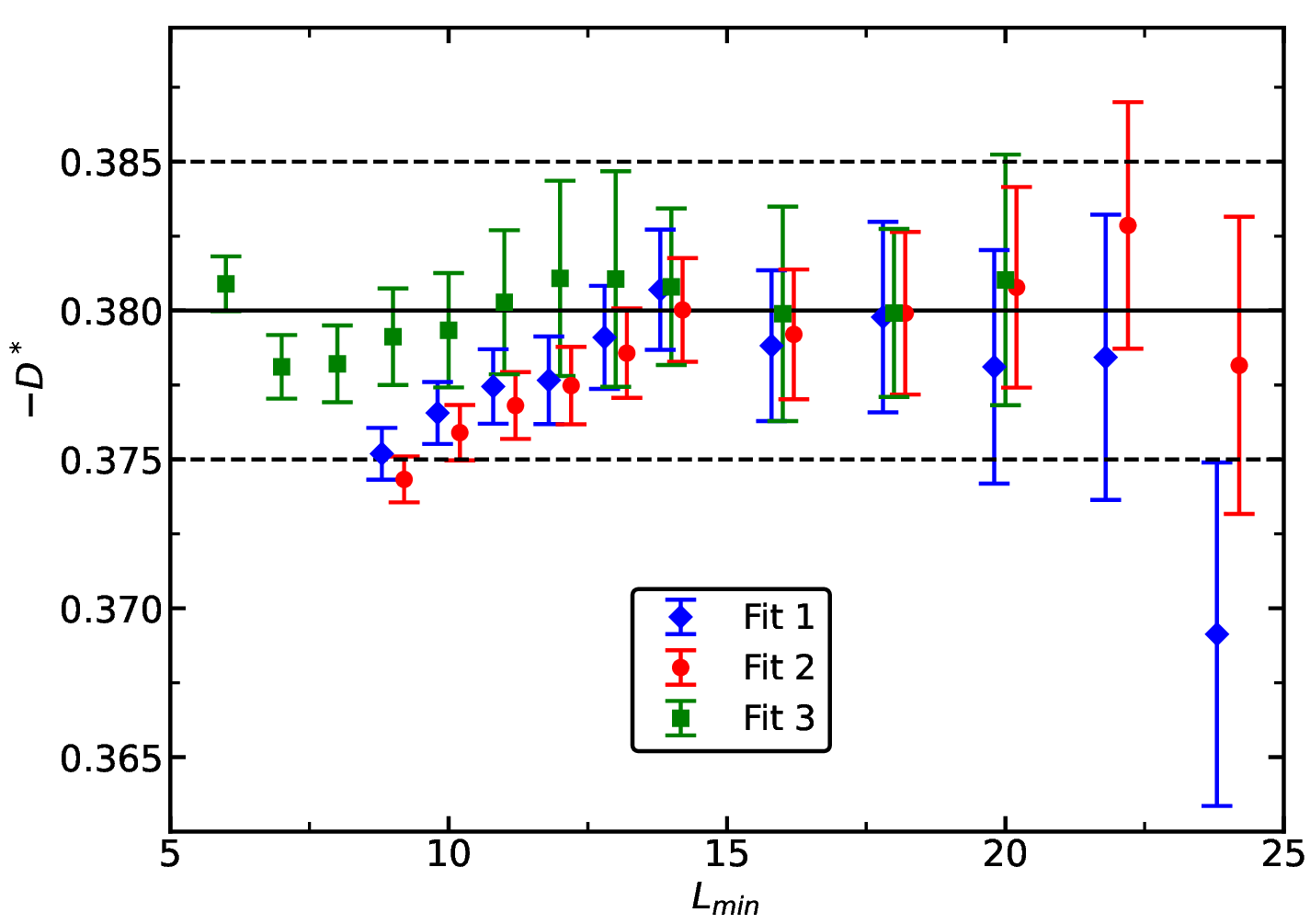}
\caption{\label{Dsplot}
Estimates of $-D^*$ plotted versus the minimal lattice size $L_{min}$
taken into account in the fit.
The numerical estimates of $-D^*$ are obtained from the fits 1, 2 and 3, which
are discussed in the text. The solid black line gives our final estimate
of $-D^*$, while the dashed lines indicate the error bar. 
Note that the values on the $x$-axis are slightly shifted to reduce overlap
of the symbols.
}
\end{center}
\end{figure}

In Fig. \ref{Dsplot} we plot the estimates of $-D^*$ obtained by the three 
different fits as a function of $L_{min}$. We quote as final result
\begin{equation}
\label{Dstar}
 D^* = -0.380(5) \;.
\end{equation}
The central value and the error bar are chosen such that for 
$10 \le L_{min} \le 18$ the results of the three fits, including their error
bars, are covered.

\begin{figure}
\begin{center}
\includegraphics[width=14.5cm]{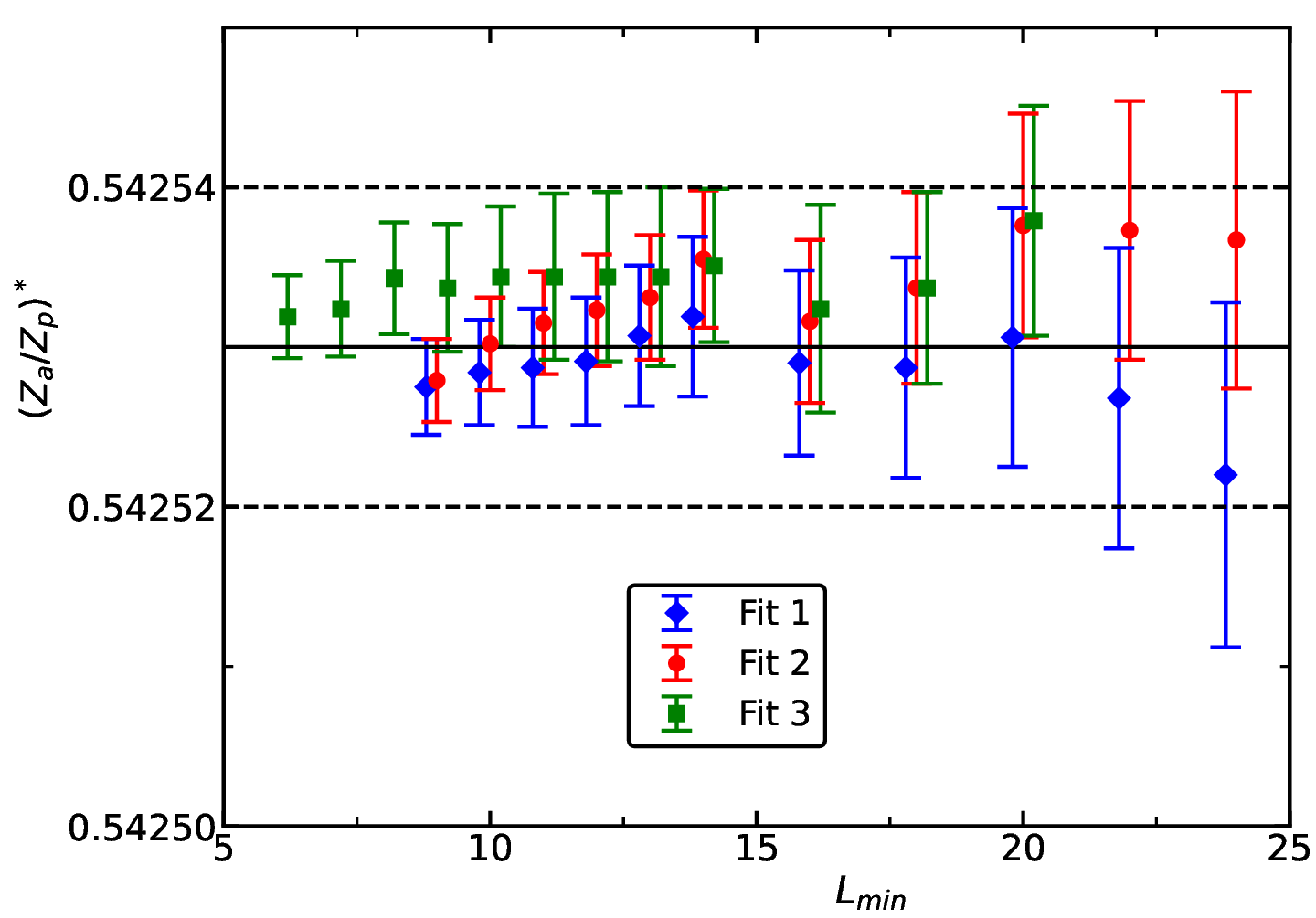}
\caption{\label{ZaZpS}
Numerical estimates of $(Z_a/Z_p)^*$ obtained from the fits 1, 2 and 3, which 
are discussed in the text. These estimates are plotted versus the minimal 
lattice size $L_{min}$ taken into account in the fit. 
The solid black line gives our final estimate
of $(Z_a/Z_p)^*$, while the dashed lines indicate the error bar.
Note that the values on the $x$-axis are slightly shifted to reduce overlap
of the symbols.
}
\end{center}
\end{figure}

In Fig. \ref{ZaZpS} we plot our estimates of $(Z_a/Z_p)^*$ obtained by using 
fits 
1, 2, and 3. The value of our final result is determined by fit 1 for 
$L_{min} =11$
up to $20$. The error bar is chosen such that up to $L_{min}=18$ the results,
including their error bars, of all three fits are covered. We quote
\begin{equation}
(Z_a/Z_p)^*=0.54253(1) \;\;.
\end{equation}
Performing a similar analysis we arrive at $U_4^*=1.60359(4)$, 
$U_6^*=3.10535(10)$, and $(\xi_{2nd}/L)^* = 0.64312(1)$. These numbers 
can be compared with 
$(Z_a/Z_p)^*=0.5425(1)$, $U_4^* = 1.6036(1)$, $U_6^*=3.1053(5)$, and 
$(\xi_{2nd}/L)^*=0.6431(1)$, which  were obtained in \cite{MHcritical}.
Note that in the analysis of \cite{MHcritical} we assumed that there is a
correction with the exponent $\omega'=1.67(11)$, ref. \cite{NewmanRiedel}, 
which leads to an increase of the systematic error compared with
the hypothesis that there is no such correction. Furthermore, the
statistics in the present study is considerably larger than that of 
\cite{MHcritical}. 
From a finite size scaling study of the Ising model on the simple cubic 
lattice the authors of ref. \cite{Landau18}
get $U_4^*=1.60356(15)$. Note that the authors use a different definition of 
$U_4$. We have converted their numerical result correspondingly. As an 
example of many older results we quote $U_4^*=1.6044(10)$ \cite{BlLuHe95}.
Note the authors of \cite{BlLuHe95} performed a joint analysis of several
different models that are supposed to share the three dimensional Ising 
universality class.

Note that the fixed point values $R^*$ of dimensionless quantities depend 
on the universality class. Furthermore one should notice that $R^*$ depends
on the global geometry of the system. The numbers quoted here are only valid
for the torus geometry with $L_0=L_1=L_2=L$.

Finally, in table \ref{Kctab}  we give our estimates of the critical value $K_c$
of the coupling $K$. The error is estimated in a similar fashion as for the 
quantities discussed above.
\begin{table}
\caption{\sl \label{Kctab}
Results for the critical coupling $K_c$ for different values of $D$
at $q_3=0.129$. For a discussion see the text.}
\begin{center}
\begin{tabular}{ll}
\hline
\mc{1}{c}{$D$} &  \mc{1}{c}{$K_c$} \\
\hline
 $-$0.3   &  0.234765504(20) \\
 $-$0.35  &  0.232071588(15) \\
 $-$0.38  &  0.230514310(10) \\
 $-$0.4   &  0.229500032(12) \\
 $-$0.42  &  0.228504501(14) \\
 $-$0.46  &  0.226568459(20) \\
\hline
\end{tabular}
\end{center}
\end{table}

\subsection{$U_4$ and $U_6$ at fixed values of $Z_a/Z_p$ or $\xi_{2nd}/L$}
\label{secbar}
In order to get an estimate of $\omega$ and a check of the results 
of the previous section, we analyze, similar to previous work, see
for example \cite{MHcritical,myClock,myIco}, 
$U_4$ and $U_6$ at fixed values of $Z_a/Z_p$ or $\xi_{2nd}/L$.   As discussed 
for example in \cite{myClock}, it is advantageous to fix $Z_a/Z_p$ and  
$\xi_{2nd}/L$ to good 
approximations of their fixed point values, respectively.  Here we take
$Z_a/Z_p = 0.54253$ and $\xi_{2nd}/L= 0.64312$. The quantities behave as
\begin{equation}
\label{baransatz}
 \bar{U}_4 = \bar{U}_4^* + \bar{b}(D) L^{-\omega} + 
               \bar{b}_2 [\bar{b}(D) L^{-\omega}]^2
 + ... + \bar{c}(D) L^{-\epsilon} + ... \.
\end{equation}
The bar on top of the quantities refers to the fact that the quantity is taken 
at either $Z_a/Z_p = 0.54253$ or $\xi_{2nd}/L= 0.64312$. This means that
we evaluate for each lattice size $L$ the value of $K$, where $Z_a/Z_p$ or
$\xi_{2nd}/L$ assumes the desired value. Then $U_4$ and $U_6$ are evaluated
at this particular value of $K$. 
For a more detailed discussion of eq.~(\ref{baransatz}), 
see section III of ref. \cite{myClock} and
references therein.  In our Ans\"atze we did not use the term 
$\bar{b}_2 [\bar{b}(D) L^{-\omega}]^2$, since $\bar{b}(D)$ is small for the 
values of $D$ that we consider. In this section, $\omega$ is a free 
paramter of the fits.
The term $\bar{c}(D) L^{-\epsilon}$ represents subleading corrections.

In the case of $U_4$ the leading one is $c(D) L^{-2+\eta}$ due to the 
analytic background of the magnetic susceptibility. In addition, in the 
case of $\xi_{2nd}/L$, we expect a correction with the exponent
$\epsilon_2=2$. The 
correction $L^{-\omega_{NR}}$ should be highly suppressed in our case.

We consider the two Ans\"atze 
\begin{equation}
\label{a1}
 \bar{U}_4 = \bar{U}_4^* + \bar{b}(D) L^{-\omega} +
 \bar{c}_1(D) L^{-\epsilon_1} 
\end{equation}
and
\begin{equation}
\label{a2}
 \bar{U}_4 = \bar{U}_4^* + \bar{b}(D) L^{-\omega} +
 \bar{c}_1(D) L^{-\epsilon_1} + \bar{c}_2(D) L^{-\epsilon_2} \;\;.
\end{equation}
We parameterized $\bar{b}(D)$ by
\begin{equation} 
 \bar{b}(D)= \bar{b}_1 (D-D^*) + \frac{1}{2} \bar{b}_2 (D-D^*)^2 \;,
\end{equation}
where the free parameters are $D^*$, $\bar{b}_1$, and $\bar{b}_2$. 
An advantage of this
parameterization is that $D^*$ is a direct outcome of the fit. Since the 
values of $D$ are contained in a narrow interval, 
we assumed $\bar{c}_1(D)$ and $\bar{c}_2(D)$
to be constant in the fit. 

First we analyzed our data for $U_4$ at $Z_a/Z_p = 0.54253$. Here we only used
Ansatz~(\ref{a1}), with $\epsilon_1=2-\eta$ as subleading correction 
exponent. Fitting data for all values of $D$, we get $\chi^2$/d.o.f. 
$=4.58$, $1.30$, $1.13$, and $1.08$ for $L_{min}=6$, $7$, $8$, and $9$,
respectively. Going to larger $L_{min}$, $\chi^2$/d.o.f. remains slightly 
larger than one.

Next, we analyzed our data for $U_4$ at $\xi_{2nd}/L= 0.64312$ by using
the Ansatz~(\ref{a1}) with $\epsilon_1=2-\eta$. Fitting data for all values of 
$D$, we get $\chi^2$/d.o.f. $=2.72$, $1.90$, $1.50$, $1.31$, $1.11$, 
and $1.06$, for $L_{min}=6$, $7$, $8$, $9$, $10$, and $11$, respectively.
For $L_{min} \ge 12$, $\chi^2$/d.o.f. drops slightly below one.

Since for fixing $\xi_{2nd}/L= 0.64312$  the $\chi^2/$d.o.f. decreases 
more slowly with increasing $L_{min}$ at small $L_{min}$ than for 
fixing $Z_a/Z_p = 0.54253$ and also motivated by the behavior of the results
for $D^*$, we analyzed our data for $U_4$ at $\xi_{2nd}/L= 0.64312$ 
in addition by using the Ansatz~(\ref{a2}) with $\epsilon_1=2-\eta$ and 
$\epsilon_2=2$. Here we find $\chi^2$/d.o.f. 
$=0.938$ already for $L_{min}=6$. For larger values 
of $L_{min}$ it stays below one.

\begin{figure}
\begin{center}
\includegraphics[width=14.5cm]{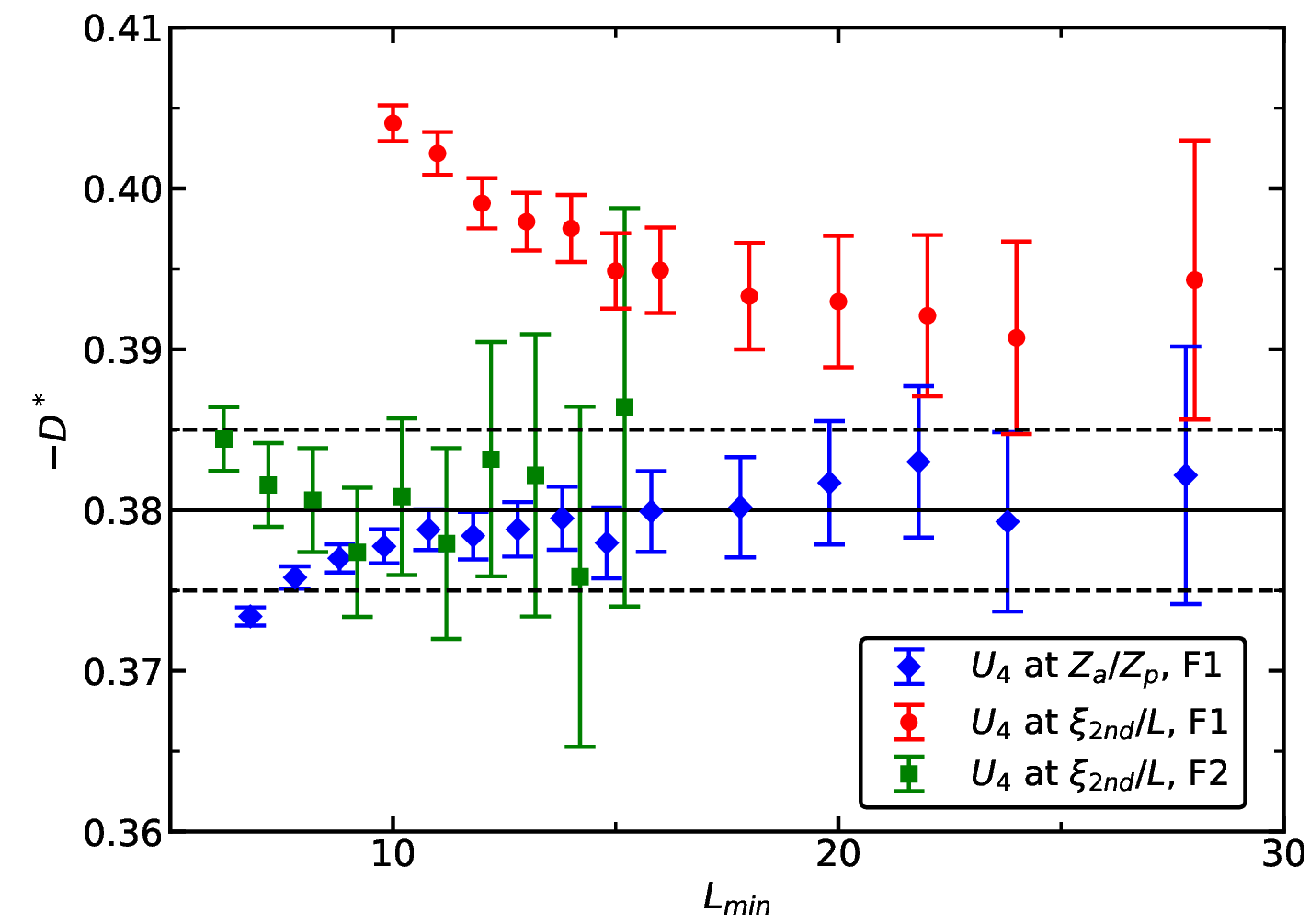}
\caption{\label{Dsfixplot}
We plot estimates of $-D^*$ obtained by fitting $U_4$ at $Z_a/Z_p = 0.54253$
and $\xi_{2nd}/L= 0.64312$  by using the Ansatz~(\ref{a1}) versus the 
minimal lattice size $L_{min}$ taken into account in the fit.  In the legend
the Ansatz~(\ref{a1}) is indicated by F1. In the case of 
fixing $\xi_{2nd}/L= 0.64312$  we fitted in addition by using 
the Ansatz~(\ref{a2}), which is indicated by F2. The 
solid and the dashed lines give the final result of the previous section  
and the corresponding error bar.
Note that the values on the $x$-axis are slightly shifted to reduce overlap
of the symbols.
}
\end{center}
\end{figure}
In Fig.  \ref{Dsfixplot} we give our results for $D^*$ obtained by using these
three different fits. In the case of fixing $Z_a/Z_p = 0.54253$ the estimate
is consistent with the one of the previous section, starting from $L_{min}=8$.
The situation is quite different for fixing $\xi_{2nd}/L= 0.64312$ and
Ansatz~(\ref{a1}). For small $L_{min}$ the estimate of $-D^*$
is too large compared with the one of the previous 
section and only slowly decreases with increasing $L_{min}$. In contrast, 
using the Ansatz~(\ref{a2}), we see consistent results, starting from very
small $L_{min}$. We conclude that the analysis presented here, confirms
the final estimate of $D^*$, eq.~(\ref{Dstar}), given above.

\begin{figure}
\begin{center}
\includegraphics[width=14.5cm]{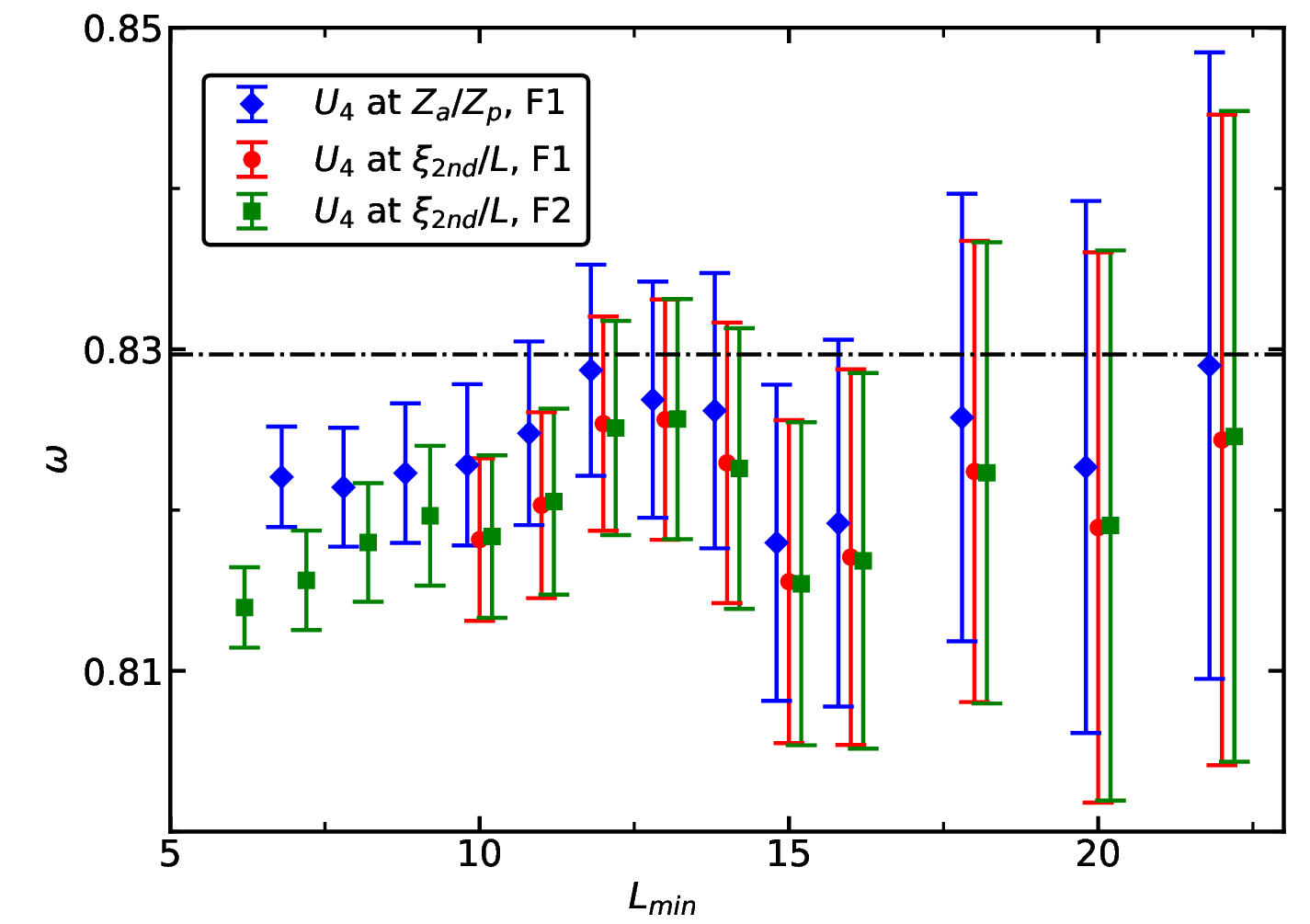}
\caption{\label{Omplot}
We plot estimates of $\omega$ obtained  
by fitting $U_4$ at $Z_a/Z_p = 0.54253$
and $\xi_{2nd}/L= 0.64312$  by using the Ansatz~(\ref{a1}) versus
the minimal lattice size $L_{min}$ taken into account in the fit.  
In the legend the Ansatz~(\ref{a1}) is indicated by F1. In the case of
fixing $\xi_{2nd}/L= 0.64312$ we fitted in addition by using
the Ansatz~(\ref{a2}), which is indicated by F2. 
Note that the values on the $x$-axis are slightly shifted to reduce overlap
of the symbols.  The
dash-dotted line gives the result of ref. \cite{Simmons-Duffin:2016wlq}.
}
\end{center}
\end{figure}

Next, in Fig. \ref{Omplot}, we plot estimates of $\omega$ obtained by these
three fits.  In contrast to $D^*$, there is very little difference between 
the results of the different fits. In  Fig. \ref{Omplot}, we give the estimate 
$\omega=0.82968(23)$ of ref. \cite{Simmons-Duffin:2016wlq} for comparison. 
Our data are certainly consistent with this estimate. As our final estimate
we might quote $\omega=0.825(20)$. This is less precise than $\omega=0.832(6)$
given in ref. \cite{MHcritical}.  Note that the present study was not designed
for an accurate estimate of $\omega$. To this end, a larger range of $D$ 
is needed.

\begin{figure}
\begin{center}
\includegraphics[width=14.5cm]{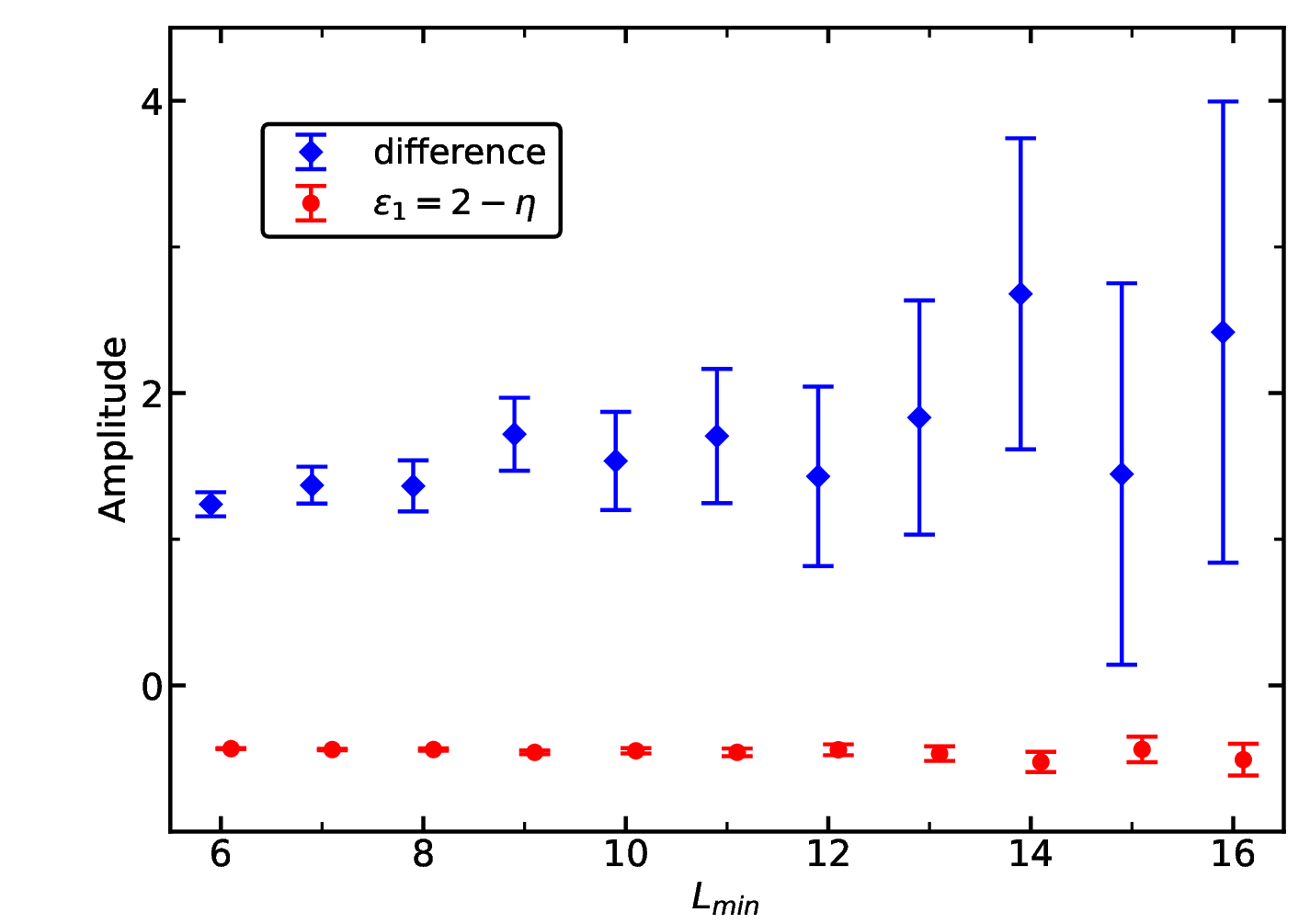}
\caption{\label{ampliplot}
We plot estimates of amplitudes of subleading corrections obtained by
fitting $U_4$ at $\xi_{2nd}/L= 0.64312$ by using the Ansatz~(\ref{a2Y})
versus the minimal lattice size $L_{min}$ taken into account in the fit.
We give the amplitude $\bar{c}$ of $L^{-\epsilon_1}$ and $\bar{d}$ of 
the difference $L^{-\epsilon_1} -L^{-\epsilon_2}$.
In the legend we give either the value of the correction exponent $\epsilon_1$
or ``difference''.
Note that the values on the $x$-axis are slightly shifted to reduce overlap
of the symbols.  
}
\end{center}
\end{figure}

In order to understand better the interplay of the two corrections, 
we have rewritten the Ansatz~(\ref{a2}) in the form
\begin{equation}
\label{a2Y}
 \bar{U}_4 = \bar{U}_4^* + \bar{b}(D) L^{-\omega} +
 \bar{c}(D) L^{-\epsilon_1} + \bar{d}(D) 
(L^{-\epsilon_1} -L^{-\epsilon_2})  \;\;.
\end{equation}
Here we fitted with $\omega=0.82968$ fixed and $\bar{c}$ and $\bar{d}$ 
not depending 
on $D$. Data for $U_4$ at $\xi_{2nd}/L= 0.64312$ for $D=-0.35$, $-0.38$, 
$-0.4$, and $-0.42$ are included 
in the fit.  The estimates of the amplitudes $\bar{c}$ and $\bar{d}$ are 
plotted in Fig. \ref{ampliplot}. The results clearly indicate that
there are two different corrections with exponents $\epsilon \approx 2$.
Furthermore, the fact that $|\bar{d}|$ being clearly larger than 
$| \bar{c}|$ shows that at least for the lattice sizes $L$ considered
here, the corrections numerically cancel to a considerable extent.
Just to given an idea, for example $10^{-\eta}= 0.9198...$ or 
$100^{-\eta}= 0.8460...$, using the CB estimate $\eta=0.0362978$
\cite{Kos:2016ysd,Simmons-Duffin:2016wlq}. 
This fact might explain the behavior of the results for $-D^*$ obtained 
by fitting $U_4$ at $\xi_{2nd}/L= 0.64312$ with the Ansatz~(\ref{a1}).

We also analyzed  $U_6$ at fixed values of $Z_a/Z_p$ or $\xi_{2nd}/L$.
Since the results are very similar to those for $U_4$, 
we abstain from a discussion. 

Finally, we have reanalyzed our data for the standard Ising model obtained 
in ref.  \cite{MHcritical}. We determined the amplitude $\bar{b}$ of the
leading correction 
in $U_4$ at $Z_a/Z_p = 0.54253$ using $\omega=0.82968$ as input. 
Combining the result $\bar{b}_{Ising} \approx -0.2$ of this analysis 
with the data obtained here for 
the derivative of the  amplitude of the leading correction with respect to
$D$ at $D=-0.38$, we conclude that at $D=-0.38$, for $q_3=0.129$, leading 
corrections to scaling are
suppressed at least by a factor of about 270 compared with the standard 
Ising model on the simple cubic lattice.

\subsection{The magnetic susceptibility}
In order to determine the critical exponent $\eta$, we analyze the 
magnetic susceptibility $\chi$ at $Z_a/Z_p= 0.54253$ or $\xi_{2nd}/L= 0.64312$.
Fixing $Z_a/Z_p$, no additional corrections with $\epsilon \approx 2$ 
are introduced. For $\xi_{2nd}/L$ fixed, the
statistical error is smaller. However the analysis of the data is more 
difficult due to subleading corrections with the exponent $\epsilon_2=2$.

In addition to the magnetic susceptibility $\bar {\chi}$ at a fixed 
value of a dimensionless quantity, we analyzed the 
improved version of it
\begin{equation}
\label{chiI}
 \bar{\chi}_{imp}= \bar{\chi} \bar{U}_4^x  \;,
\end{equation}
where the bar indicates that the quantity is taken at a fixed value of 
$Z_a/Z_p$ or $\xi_{2nd}/L$.  The exponent $x$ is tuned such that leading 
corrections to scaling are eliminated. For simplicity, we took the 
result obtained in section VII of ref. \cite{MHcritical}:  
$x=-0.66$ and $x=-0.57$ for fixing $Z_a/Z_p$ and $\xi_{2nd}/L$, 
respectively. 

\begin{figure}
\begin{center}
\includegraphics[width=14.5cm]{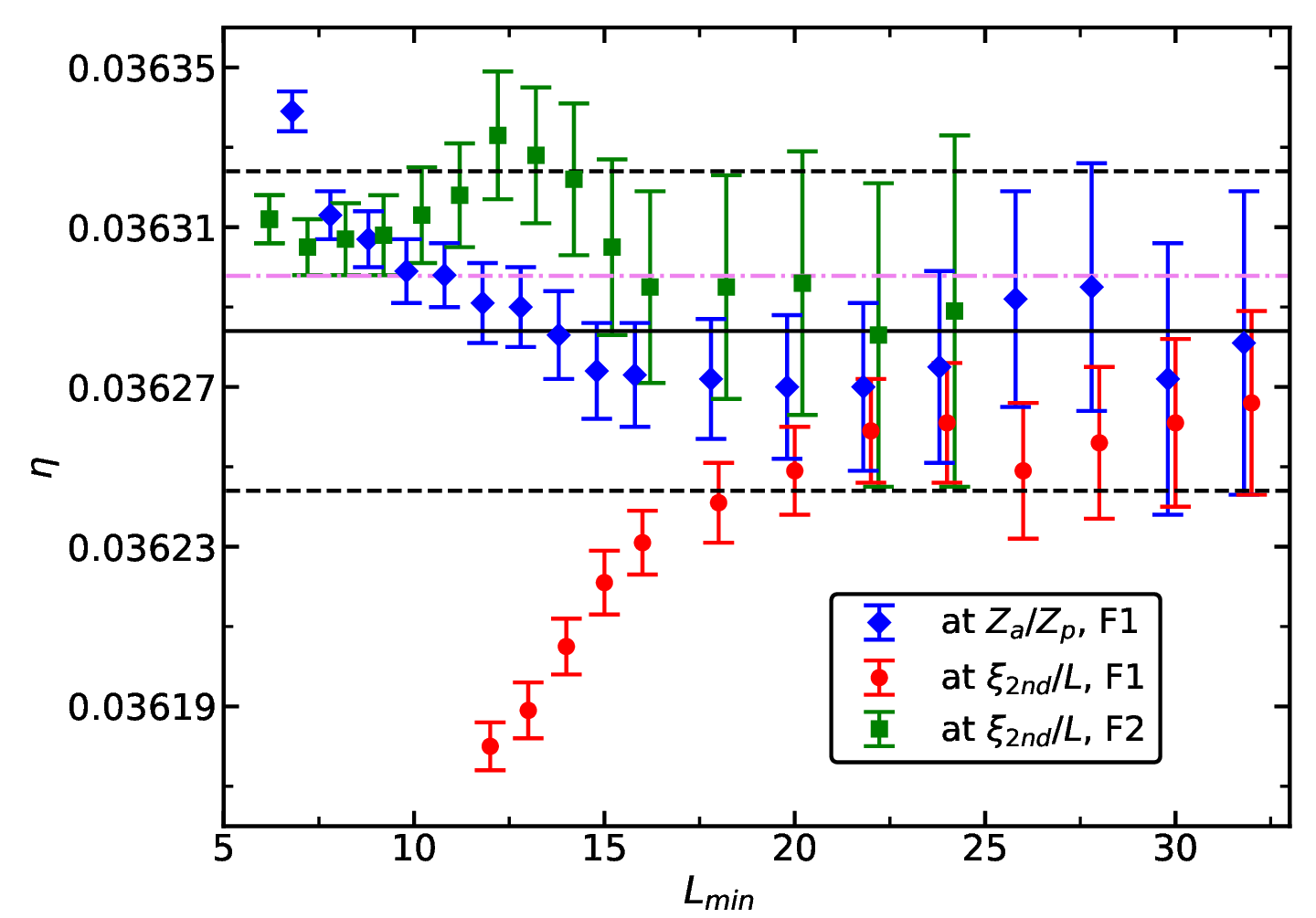}
\caption{\label{etaplot}
Numerical estimates of $\eta$ for $D=-0.38$ plotted versus the minimal 
lattice size $L_{min}$ taken into account in the fit. 
We have fitted the non-improved $\chi$ by using the Ansatz~(\ref{chi1})
for both fixing $Z_a/Z_p$ and $\xi_{2nd}/L$. These fits are denoted 
by F1 in the legend. In the case of fixing $\xi_{2nd}/L$, we give 
results obtained by using the Ansatz~(\ref{chi2}) in addition. 
These fits are denoted by F2 in the legend.
The values on the $x$-axis are slightly shifted to reduce overlap
of the symbols. 
The solid black line gives our final estimate
of $\eta$, while the dashed lines indicate the error bar.
The dash-dotted line gives
the estimate of refs. \cite{Kos:2016ysd,Simmons-Duffin:2016wlq}.
Note that the error bar of the 
CB result is by a factor of 20 smaller than the one obtained here.
}
\end{center}
\end{figure}

We fit our data for fixing $Z_a/Z_p$ and $\xi_{2nd}/L$ 
with the Ansatz
\begin{equation}
\label{chi1}
\bar{\chi}= \bar{c} L^{2-\eta} + \bar{b} \;.
\end{equation}
In the case of fixing 
$\xi_{2nd}/L$ we used in addition 
\begin{equation}
\label{chi2}
\bar{\chi}= \bar{c} L^{2-\eta} \; (1+\bar{d} L^{-2}) \;  + \bar{b} \;.
\end{equation}
Our results for $D=-0.38$ and the standard version of $\bar{\chi}$ are given
in Fig.  \ref{etaplot}. In the case of fixing $Z_a/Z_p$, $\chi^2/$d.o.f
already drops below one at $L_{min}=9$. In contrast, for fixing 
$\xi_{2nd}/L$ it drops below two at $L_{min}=16$ and remains larger than
$1.5$ even for larger $L_{min}$. Using the Ansatz~(\ref{chi2}) we get 
$\chi^2/$d.o.f. $=1.23$ for $L_{min}=6$, corresponding to $p=0.16$. 
The $\chi^2/$d.o.f. stays roughly at this level going to larger values of 
$L_{min}$. 
One might be tempted to take $\eta=0.036299(8)$ obtained from the 
fit using the Ansatz~(\ref{chi1}) of $\chi$ at $Z_a/Z_p=0.54253$ with 
$L_{min}=10$ as 
final result, where $\chi^2/$d.o.f. $=0.93$ corresponding to $p = 0.58$.

Nevertheless, as our final estimate we quote the more cautious
\begin{equation}
\label{finaleta}
 \eta = 0.036284(40) \;\;. 
\end{equation}
It is chosen such that it covers the results, including their error bars, 
obtained by fitting $\chi$ at $Z_a/Z_p=0.54253$ by using the 
Ansatz~(\ref{chi1}) up to $L_{min}=26$. In addition 
the results, including their error bars, obtained by fitting $\chi$ at
$\xi_{2nd}/L= 0.64312$ by using the Ansatz~(\ref{chi1}) for 
$L_{min}=22$ and $24$ are covered. 
In the case of fitting $\chi$ at $\xi_{2nd}/L= 0.64312$
by using the Ansatz~(\ref{chi2}) the results, including their error bars,
are covered for the majority of $L_{min}$-values with $L_{min} \le 22$.

\begin{figure}
\begin{center}
\includegraphics[width=14.5cm]{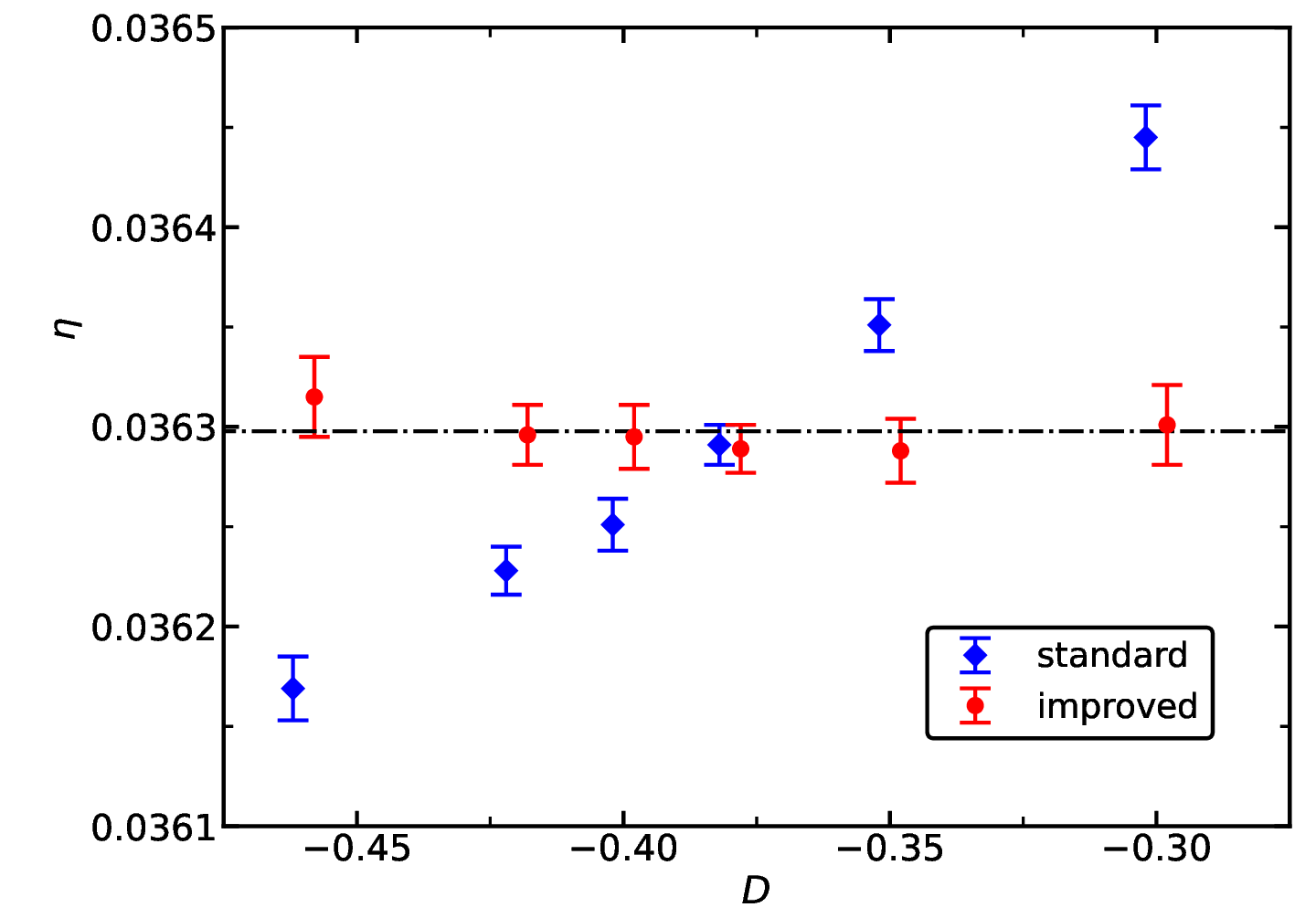}
\caption{\label{etaplotI}
Numerical estimates of $\eta$ obtained by fitting $\chi$ at 
$Z_a/Z_p= 0.54253$ by using the Ansatz~(\ref{chi1}) versus $D$. 
We compare results obtained for the standard and the improved, 
eq.~(\ref{chiI}),
version of the magnetic susceptibility. In all cases we give the estimate
obtained with $L_{min}=12$. 
The values on the $x$-axis are slightly shifted to reduce overlap
of the symbols. The dash-dotted line gives
the estimate of refs. \cite{Kos:2016ysd,Simmons-Duffin:2016wlq}.
}
\end{center}
\end{figure}

Finally we study the effect of deviations of $D$ from $D^*$. To this 
end we have fitted $\chi$ at $Z_a/Z_p= 0.54253$ and its improved version 
for all values of $D$ we simulated at by using the Ansatz~(\ref{chi1}).
Our results for $L_{min}=12$ are given in Fig.  \ref{etaplotI}. We see
that the estimate of $\eta$ obtained from $\bar{\chi}$ has a clear dependence 
on $D$. In contrast, this dependence is, within errors, eliminated for 
$\bar{\chi}_{imp}$. This finding confirms that the exponent $x$ is universal.
Note that in ref. \cite{MHcritical}, we have determined $x$ by studying 
the Blume-Capel model at $q_3=0$.  Furthermore, we see that for $D=-0.38$
the estimates obtained by fitting $\bar{\chi}$ and $\bar{\chi}_{imp}$ 
essentially
coincide. This confirms our result $D^*=-0.380(5)$, eq.~(\ref{Dstar}),
obtained above. Furthermore
no revision of our final estimate, eq.~(\ref{finaleta}), is needed.

\subsection{The exponent $\nu$}
We study the slopes of dimensionless quantities at a fixed value of 
a dimensionless quantity. Below we restrict the discussion on 
fixing $Z_a/Z_p=0.54253$, since $Z_a/Z_p$ has virtually no corrections
$\propto L^{-\epsilon}$, where $\epsilon \approx 2$.

\subsubsection{Analyzing quantity by quantity}
\begin{figure}
\begin{center}
\includegraphics[width=14.5cm]{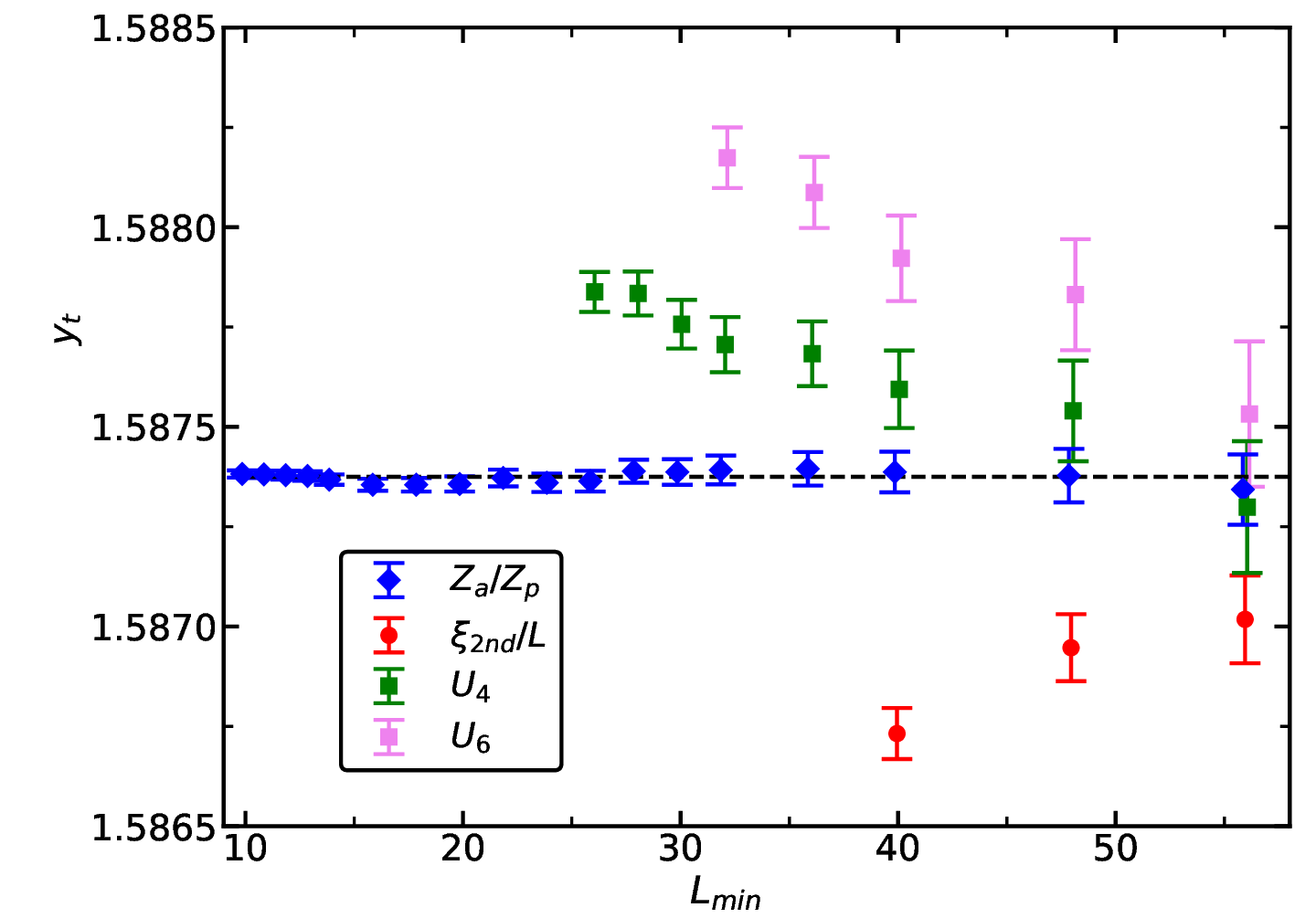}
\caption{\label{Figyt0}
Numerical estimates of $y_t$ for $D=-0.38$ obtained by fitting the slopes 
of dimensionless quantities at $Z_a/Z_p=0.54253$ by using the 
Ansatz~(\ref{Sansatz0}) versus the minimal lattice size $L_{min}$ that 
is taken into account. The dash-dotted line gives
the estimate of refs. \cite{Kos:2016ysd,Simmons-Duffin:2016wlq}. 
The legend refers to the quantity that is analyzed.
}
\end{center}
\end{figure}

In this section we have fitted the slopes of dimensionless quantities
one by one. First we have fitted the data for $D=-0.38$ with the Ansatz
\begin{equation}
\label{Sansatz0}
 \bar{S}_R = \bar{a} L^{y_t} \;,
\end{equation}
not taking into account corrections.
The estimates of $y_t$ obtained this way are given in Fig. \ref{Figyt0}.
In the figure we give only results that correspond to $p>0.01$. For 
the slope of $Z_a/Z_p$, actually already for $L_{min}=10$, we get 
$\chi^2/$d.o.f.$=0.96$ corresponding to $p=0.52$. 
Furthermore we see that for the slope of $Z_a/Z_p$ 
the estimate of $y_t$ does change little with increasing $L_{min}$.
For $L_{min}=10$ we get $y_t=1.587382(9)$, consistent with the CB estimate
$y_t=1.587375(10)$ \cite{Kos:2016ysd,Simmons-Duffin:2016wlq}. 

In the case of the other 
quantities, we see a clear dependence of the estimate on $L_{min}$. 
At least, in all cases the CB estimate is approached as $L_{min}$ increases.
This observation is consistent with the fact that only in the case of the 
slope of $Z_a/Z_p$ we do not expect corrections $\propto L^{-\epsilon}$ with
$\epsilon \approx 2$. 

\begin{figure}
\begin{center}
\includegraphics[width=14.5cm]{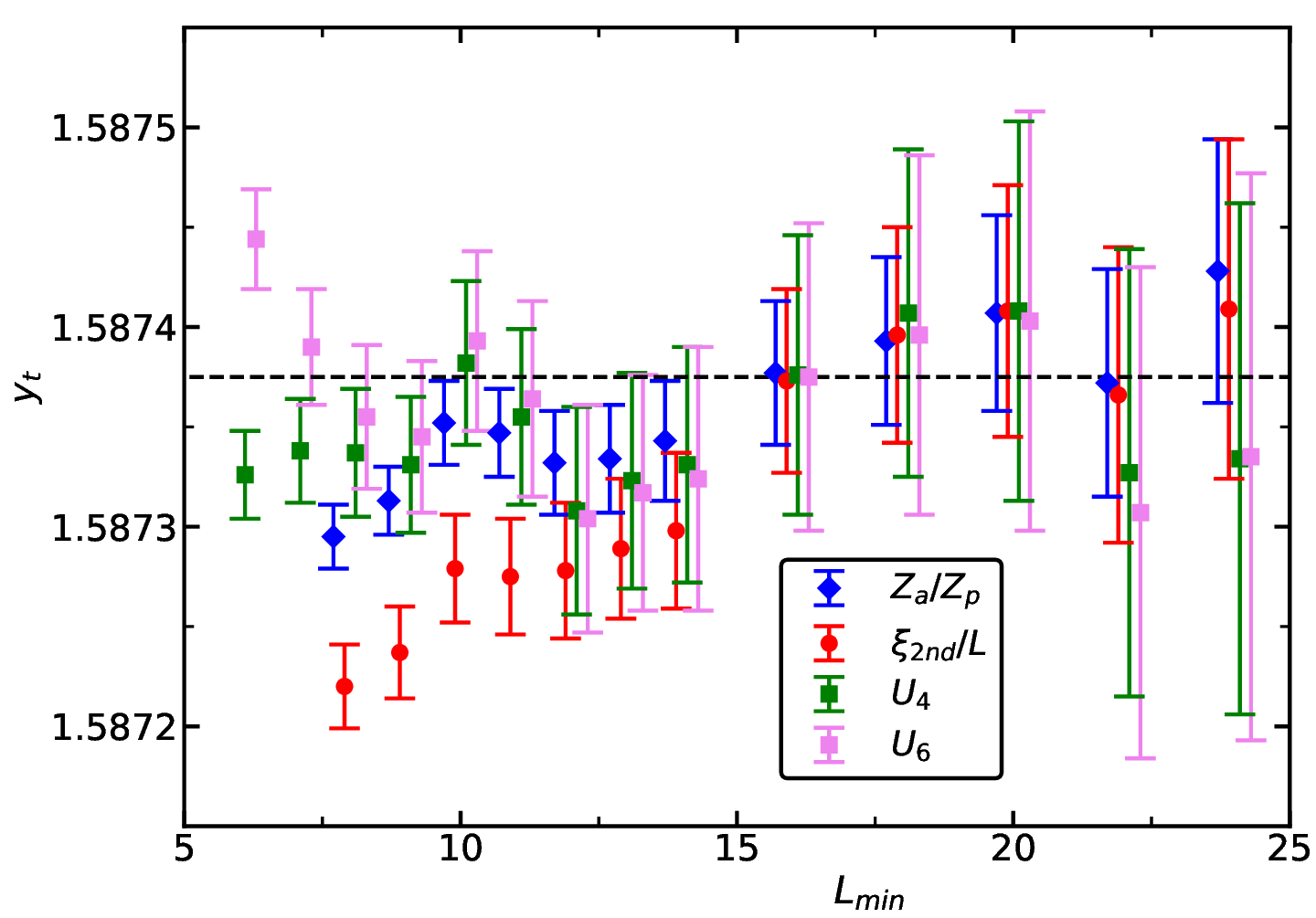}
\caption{\label{Figyt1}
Numerical estimates of $y_t$ for $D=-0.38$ obtained by fitting the slopes 
of dimensionless quantities with the Ansatz~(\ref{Sansatz1}) versus 
the minimal lattice size $L_{min}$ that is taken into account.
The dash-dotted line gives the estimate of refs. 
\cite{Kos:2016ysd,Simmons-Duffin:2016wlq}.
The legend refers to the quantity that is analyzed in the fit.
}
\end{center}
\end{figure}

Next, in Fig. \ref{Figyt1} we give estimates of $y_t$ obtained by fitting 
with the Ansatz
\begin{equation}
\label{Sansatz1}
 \bar{S}_R = \bar{a} L^{y_t}  \; (1+ \bar{c} L^{-\epsilon}) \;\;, 
\end{equation}
where $\epsilon=2 -\eta$. Here $p > 0.01$ is reached for much smaller 
$L_{min}$ than for Ansatz~(\ref{Sansatz0}). Furthermore, the estimates 
obtained by analyzing the different slopes are close to each other starting 
from small values of $L_{min}$.  For example for $L_{min}=14$ the preliminary
result
$y_t=1.58733(7)$ corresponding to $\nu=0.629989(28)$ covers the estimates
obtained by analyzing the four different quantities. It is fully consistent 
with the estimate of refs. \cite{Kos:2016ysd,Simmons-Duffin:2016wlq}.
Note that for $L_{min}=14$ we get
$\chi^2/$d.o.f.$=0.96$, $1.34$, $1.18$, and $1.26$
corresponding to $p=0.51$, $0.12$, $0.25$, and $0.17$ for the 
slopes of $Z_a/Z_p$, $\xi_{2nd}$, $U_4$, and $U_6$, respectively.

Next we focus on the slope of $Z_a/Z_p$. It should not contain a 
correction related to the analytic background and the 
contribution $\propto L^{-\omega_{NR}}$ should be very small. One expects
a correction with the correction exponent $y_t + \omega$. For a discussion 
see for example sec. III of ref. \cite{myClock}.
It turns out that replacing  
$\epsilon=2-\eta$ by $\epsilon=y_t + \omega \approx 2.417055$ in the 
Ansatz~(\ref{Sansatz1}) the numerical estimates of $y_t$ change only 
slightly.
This can be explained by the fact that the amplitude of the correction 
is small. 

\begin{figure}
\begin{center}
\includegraphics[width=14.5cm]{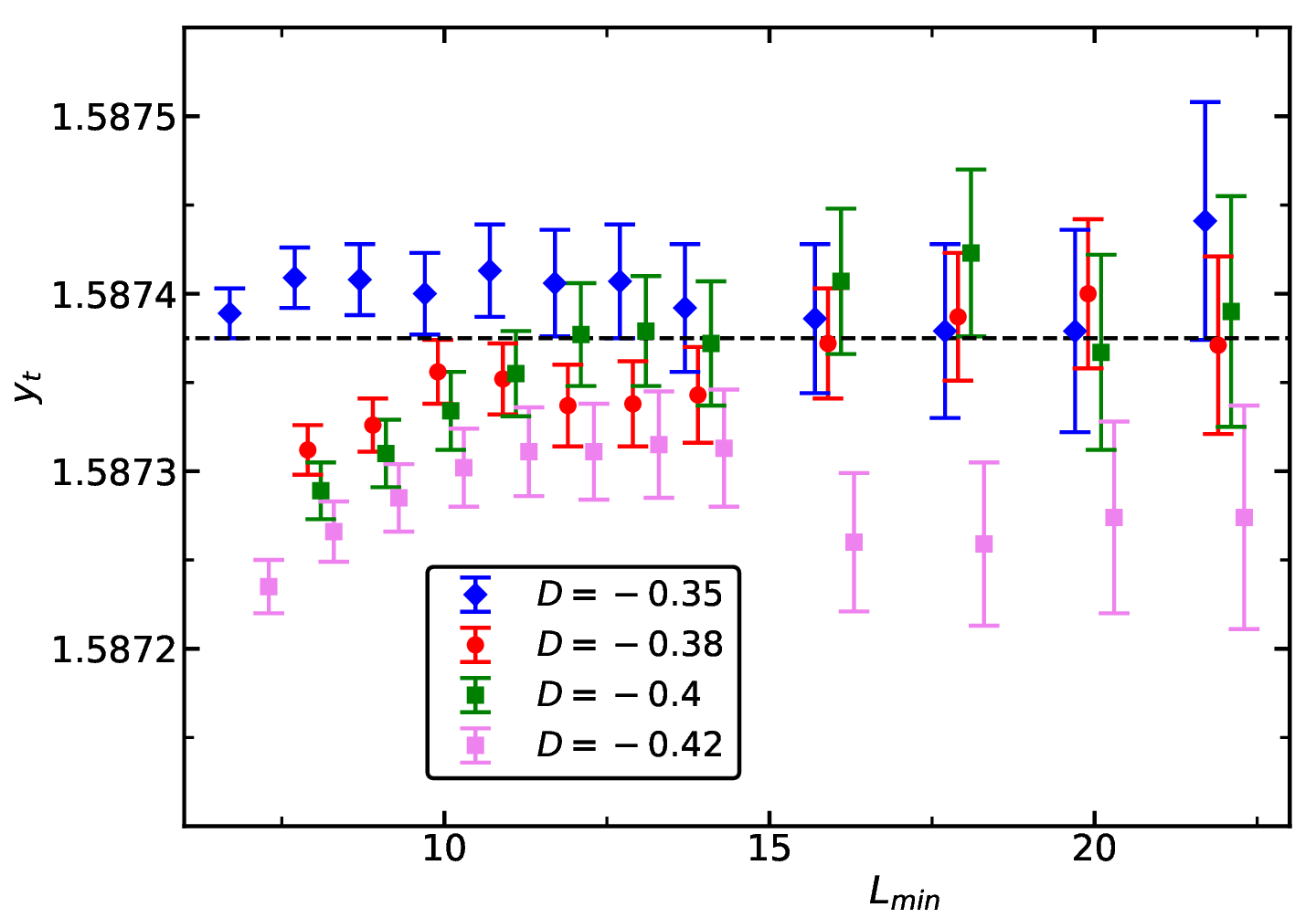}
\caption{\label{Figyt3}
Numerical estimates of $y_t$ for $D=-0.35$, $-0.38$, $-0.4$, and $-0.42$
 obtained by fitting the slope of $Z_a/Z_p$ by using the
Ansatz~(\ref{Sansatz1}) with $\epsilon=y_t + \omega$ as correction 
exponent. These estimates are plotted versus the minimal lattice 
size $L_{min}$ taken into account in the fit. 
Note that the values on the $x$-axis are slightly shifted to reduce overlap
of the symbols.
The dash-dotted line gives
the estimate of ref. \cite{Kos:2016ysd,Simmons-Duffin:2016wlq}. 
}
\end{center}
\end{figure}

In Fig. \ref{Figyt3}  we give estimates of $y_t$ obtained by fitting the
slope of $Z_a/Z_p$  with the Ansatz~(\ref{Sansatz1}) and
setting the correction exponent to $\epsilon=y_t + \omega$. We give data for
$D=-0.35$, $-0.38$, $-0.4$, and $-0.42$. We see only a small variation 
of the result with $D$.  Here we take $y_t = 1.58734(4)$ corresponding to
$\nu=0.629985(16)$ as preliminary result that
covers estimates obtained for $10 \le L_{min} \le 14$ for $D=-0.38$.
Note that $\chi^2$/d.o.f.$=0.91$  corresponding to $p=0.61$ for $L_{min}=10$.

\subsubsection{Joint fit using all four dimensionless quantities}
\begin{figure}
\begin{center}
\includegraphics[width=14.5cm]{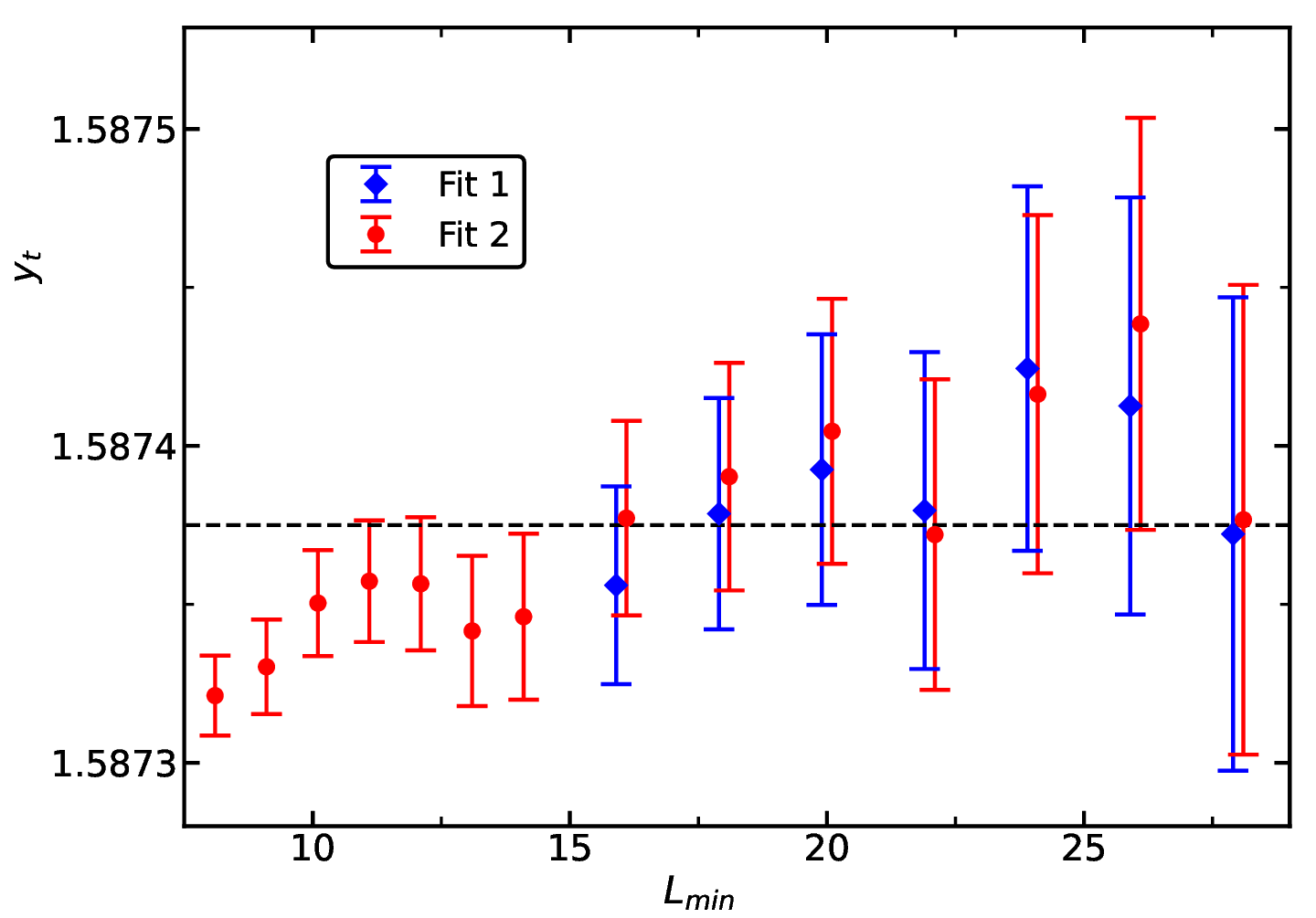}
\caption{\label{Figyt4}
Numerical estimates of $y_t$ for $D=-0.38$
obtained by fitting the slopes of all four dimensionless quantities jointly 
with the Ansatz~(\ref{Sansatz1}) or Ansatz~(\ref{Sansatz3}). 
These estimates are plotted versus the minimal lattice size $L_{min}$ that
is taken into account in the fit. In the legend,
these Ans\"atze are denoted by Fit 1 and Fit 2, respectively.
Note that the values on the $x$-axis are slightly shifted to reduce overlap
of the symbols.
}
\end{center}
\end{figure}

Finally, we performed joint fits of the slopes of all four dimensionless 
quantities using the Ansatz~(\ref{Sansatz1}) and in addition
\begin{equation}
\label{Sansatz3}
 \bar{S}_R = \bar{a} L^{y_t}  \; (1+ \bar{c}_1 L^{-\epsilon_1} + \bar{c}_2 L^{-\epsilon_2} + c_3 L^{-\epsilon_3}) \;\;,
\end{equation}
where $\epsilon_1=2 -\eta$, $\epsilon_2=2$, and $\epsilon_3=y_t + \omega$.
We set $\bar{c}_1=0$ for the slope of $Z_a/Z_p$  and $\bar{c}_2=0$ 
for the slopes of $Z_a/Z_p$, $U_4$ and $U_6$.
In Fig. \ref{Figyt4} we give the results for $y_t$ obtained by performing
these fits using our data for $D=-0.38$. It turns out that for the 
Ansatz~(\ref{Sansatz3}) an acceptable 
$\chi^2/$d.o.f. is reached for considerably smaller $L_{min}$ than for 
Ansatz~(\ref{Sansatz1}). On the other hand, the estimates obtained for 
$y_t$ are similar. As our preliminary result we quote
\begin{equation}
\label{finalyt}
 y_t = 1.58739(7) 
\end{equation}
corresponding to $\nu=0.629965(28)$.  It covers the estimates obtained 
by both Ans\"atze, eq.~(\ref{Sansatz1},\ref{Sansatz3}), 
for $16 \le L_{min} \le 22$.  Note that for $L_{min}=16$, we get 
$\chi^2/$d.o.f.$=1.19$ and $1.14$, corresponding to $p=0.11$ and $0.17$
for the Ans\"atze, eq.~(\ref{Sansatz1},\ref{Sansatz3}), respectively.

\begin{figure}
\begin{center}
\includegraphics[width=14.5cm]{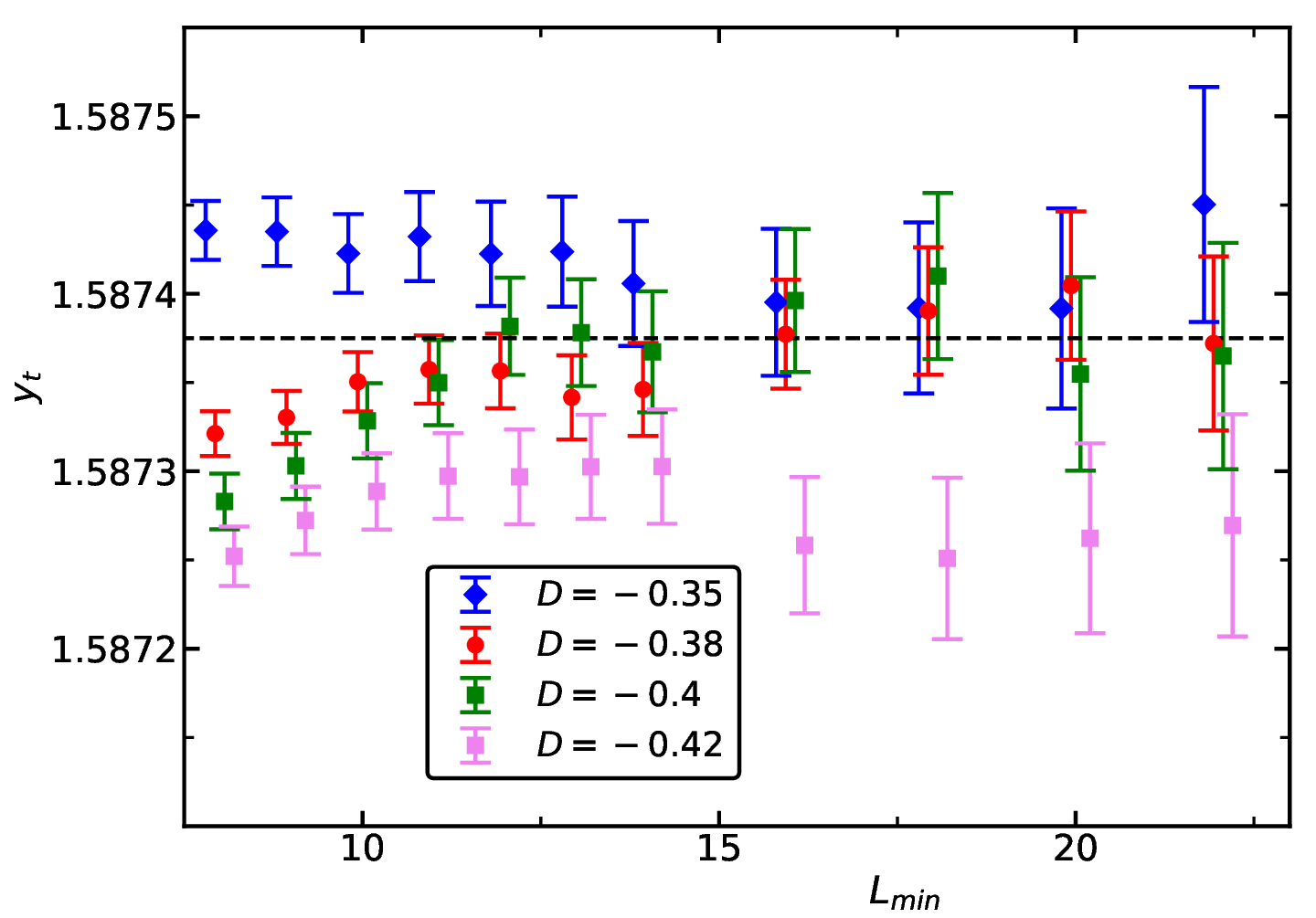}
\caption{\label{Figyt5}
Numerical estimates of $y_t$ for $D=-0.35$, $-0.38$, $-0.4$, and $-0.42$
obtained by fitting the slopes of all four dimensionless quantities jointly 
with the Ansatz~(\ref{Sansatz3}) versus the minimal lattice size $L_{min}$
taken into account.
Note that the values on the $x$-axis are slightly shifted to reduce overlap
of the symbols for different values of $D$.
}
\end{center}
\end{figure}

Finally, in Fig. \ref{Figyt5} we give results obtained by fitting the data 
for $D=-0.35$, $-0.38$, $-0.4$, and $-0.42$ using the Ansatz~(\ref{Sansatz3}).
The estimates obtained for different values of $D$ differ only by little.
Hence the estimate, eq.~(\ref{finalyt}), quoted above needs not to be 
revised.  Similar to eq.~(\ref{chiI}), improved slopes can be constructed. 
Since the dependence of our estimate of $y_t$ on $D$ is relatively small, 
we abstained from analyzing improved slopes here.

In this section we obtained three preliminary estimates of $y_t=1/\nu$ by analyzing the 
slopes of dimensionless quantities in different ways. Our final result
\begin{equation}
 \nu=0.62998(5) 
\end{equation}
covers these preliminary estimates, including their error bars.

\section{Summary and outlook}
We have studied a generalized Blume-Capel model on the simple cubic lattice. 
In addition to the nearest neighbor coupling $K_1$ there is a third nearest
neighbor coupling $K_3$. This model has been studied for example 
in ref. \cite{BlLuHe95}. Here we are aiming at the elimination of both the 
leading contribution to the spatial anisotropy and the leading correction
to scaling. This is achieved by tuning the ratio $q_3=K_3/K_1$ and the 
parameter $D$  that controls the distribution of the spin at a given site. 
For the precise definition of the reduced Hamiltonian see 
section \ref{themodel}. The values, where these corrections are eliminated
are denoted by $q_3^{iso}$ and $D^*$. It is conjectured and numerically 
confirmed that the spatial anisotropy depends little on $D$. Hence
$q_3^{iso}$ depends little on $D$.  In contrast, the leading correction 
to scaling depends on $q_3$ and $D$ in a similar strength.

In order to quantify the spatial anisotropy, we determine the correlation 
length in three different spatial directions in the high temperature phase of 
the model. We tune $q_3$ such that these three correlation lengths 
are the same. For our final estimate $q_3^{iso}=0.129(1)$,
we determine $D^*(q_3=0.129)=-0.380(5)$ by using a finite size scaling analysis
similar to that of \cite{myClock,myIco} and refs. therein.
In addition, we obtain accurate estimates of the fixed
point values of dimensionless quantities and critical temperatures. 
Furthermore, the finite size scaling analysis
provides accurate estimates of the critical exponents $\nu$ and $\eta$.

In table \ref{tabexponents} we compare these with selected results obtained by 
using different methods. For a more complete summary of  theoretical 
results given in the literature see tables 3, 4, 5, and 6  of 
ref. \cite{PeVi02} and the references given in table \ref{tabexponents}
for more recent work. A summary of experimental estimates is given in table 7
of ref. \cite{PeVi02}. In general, we find a good  agreement of the 
results obtained by the different methods. In the cases 
\cite{DengBloete03,KoPa17}, where the estimates
are slightly out of the error bars, it is plausible that these were 
underestimated, rather than that there is a fundamental problem. 
In ref. \cite{GoTr20} the Blume-Capel model at $D=0.655$ at the 
critical point has been studied with slab geometry and Dirichlet boundary
conditions. From the behavior of the magnetization as a function of the 
distance from the boundary,
compared to results of a geometric theory of bounded critical phenomena 
extending local conformal invariance to $d>2$ theories,
the authors obtain $\eta=0.036284(16)$. One should 
note that the uncertainty of $D^*$ and of the critical coupling 
$K_c$ is not taken into account in the error bar that is quoted.

\begin{table}
\caption{\sl \label{tabexponents}
Selected theoretical results for critical exponents for the three-dimensional 
Ising universality class taken from the literature. In the first column 
we indicate the method that has been used. It follows the year of the 
publication and the reference. The most accurate results
are provided by the conformal bootstrap (CB) method.  Next we give
results obtained by studies of lattice models, where either high 
temperature series expansions (HT) or Monte Carlo (MC) simulations 
are used. In some of the studies a number of different model have been
studied. This is indicated by "var". In others, the study is either restricted 
to the Blume-Capel (BC) or the Ising model.  Finally, we give two recent 
studies using the $\epsilon$-expansion and the functional renormalization 
group (FRG) method.
}
\begin{center}
\begin{tabular}{ccclll}
\hline
\mc{1}{c}{Method} &  \mc{1}{c}{Year}  &  \mc{1}{c}{ref.} &
\mc{1}{c}{$\nu$}  &  \mc{1}{c}{$\eta$} & \mc{1}{c}{$\omega$} \\
\hline
 CB        & 2016 &\cite{Kos:2016ysd,Simmons-Duffin:2016wlq} 
                        &0.6299709(40)& 0.0362978(20) & 0.82968(23) \\
 HT, var   & 2002 &\cite{pisaseries} & 0.63012(16) & 0.03639(15) & 0.825(50) \\
 MC, var   & 2003 &\cite{DengBloete03}&0.63020(12)& 0.0368(2) & 0.821(5) \\
 MC, BC    & 2010 &\cite{MHcritical}&0.63002(10) & 0.03627(10)& 0.832(6) \\
 MC, Ising & 2018 &\cite{Landau18}  &0.629912(86)& 0.03610(45)&   \\
 MC, BC, iso& 2021 & present work    & 0.62998(5) & 0.036284(40)& 0.825(20) \\  
$\epsilon$-exp.& 2017 & \cite{KoPa17} & 0.6292(5) &  0.0362(6)  & 0.820(7) \\
FRG  & 2020 & \cite{DePo20} & 0.63012(16) & 0.0361(11) & 0.832(14) \\
\hline
\end{tabular}
\end{center}
\end{table}
The results of the present study are fully consistent with those of the 
CB method. Our error bars are by a factor of 12.5 and 20 larger than 
those of the CB method for the exponents $\nu$ and $\eta$, respectively.
Nethertheless one should regard the present study as valuable consistency
check, since the approaches are complementary.

The present study gives strong support to the fact that only the breaking 
of the spatial isotropy by the lattice gives rise to a scaling field 
associated with a correction exponent $ \approx 2$.  In particular 
$\omega' = 1.67(11)$ obtained by the scaling field method \cite{NewmanRiedel}  
seems to be an artifact of the method.  In addition  
to these corrections, corrections that are intrinsic to the observable
need to be taken into account. For example the analytic background in the
magnetic susceptibility.

In the model studied here, corrections to scaling are highly suppressed. 
Compared with the standard Ising model on the simple cubic lattice, the
leading correction to isotropy is reduced by at least a
factor of $180$ and the leading correction to scaling at least by a factor of 
$270$.  Still the model is relatively simple to implement and 
can be efficiently simulated.  Hence it might be the model of choice to study
universal properties of the Ising universality class. For example 
interfacial properties, boundary critical phenomena or dynamics.

Unfortunately the idea of the present work can not be directly adopted for 
$O(N)$-symmetric models with $N>1$ as we discuss in Appendix \ref{XYnnn}.

\section{Acknowledgement}
This work was supported by the Deutsche Forschungsgemeinschaft (DFG) under the grants
No HA 3150/5-1 and HA 3150/5-2.

\appendix

\section{Random number generators}
\label{append}
Motivated by the discussion, ref. \cite{Vigna19} and refs. therein, on the 
reliability of the Mersenne Twister
algorithm \cite{twister} we performed some runs with other generators to check
the consistency of the results obtained by different generators.  

In particular, we used the double precision version of L\"uscher's RANLUX 
generator 
\cite{ranlux} at the highest luxury level. Here we made use of L\"uscher's
most recent implementation ranlux-3.4 taken from \cite{ranluxWWW}. 
Note that there are a number of 
alternative implementation. Just search the internet with your favorite 
search engine.

Furthermore, we used a generator that is based on the KISS generator proposed
by George Marsaglia. It combines three different generators as
\begin{equation}
\label{Mars}
 r = (r_1 + r_2 + r_3) \;\; \mbox{mod} \; 2^{64} \;\; ,
\end{equation}
where $r_1, r_2, r_3 \in \{0,1,2, ..., 2^{64}-1\}$ are generated by three
different, relatively simple generators.
The rough idea of such a combination is that
the generators compensate each others weaknesses. For a critical discussion 
of the KISS generator see ref. \cite{rose}. Our starting point is the 64 bit 
implementation given in the German version of \cite{KISS_wiki}.

Here, we replaced the generators $r_1$ and $r_2$ by ones that are of better 
quality than those used in Marsaglia's original generator. For $r_1$ we used
\verb$ xoshiro256+ $ taken from \cite{VignaWWW}.  For a discussion of the 
generator see \cite{ViBl18}. As second generator we used
a 96 bit linear congruential generator with the multiplier and the
increment $a=c=$\verb+0xc580cadd754f7336d2eaa27d + and the modulus $m=2^{96}$ 
suggested by O'Neill \cite{ONeill_minimal}. In this case we used our own 
implementation. In eq.~(\ref{Mars}) we use the upper $64$ bits.

Note that in the context of our simulations, the importance of the quality
of the random bits is decreasing from high to low, since the random 
numbers, normalized to the interval $[0,1)$, are used for comparisons with 
double precision floating point numbers.
It is plausible that both the generators $r_1$ and $r_2$ would 
do the job on their own, which we however did not check here.  We performed 
a few basic tests of the combined generator. In particular, we did run the 
bigcrush test \cite{TestU01} several times, with different initializations 
of the generator, on the upper, the middle and lower 
32 bits of the generator. These tests were passed.

The choice of the particular generator discussed here is essentially ad hoc
and unfortunately not based on deep insight.
While we are confident that the generator is a good choice for our purpose, we
do not recommend it for general use, since there are certainly better tested
and motivated generators that consume less CPU time.

For $D=-0.38$ we have simulated the lattice sizes $L=12$ and $120$ close
to criticality with 
roughly equal statistics using the three choices of the random number 
generator. We get consistent results for the three different choices.

The bulk of the simulations has been performed either by using 
the SFMT or the modified KISS generator. Which generator
was used is essentially determined by the history of our simulations. 
At a certain stage we switched from the SFMT to the modified 
KISS generator. As a result, the simulations for $D=-0.3$, $-0.35$, $-0.4$, 
were mainly performed by using the SFMT generator, while 
those for $D=-42$ and $-0.46$ were mainly performed by using the modified KISS
generator. In the case of $D=-0.38$ both generators have been used on 
roughly the same footing. The fact that fits that include both sets of 
simulations give reasonable $p$-values, gives us further assurance that
there is nothing terribly wrong with the generators that we have used.

To give the reader an impression on the relative performance of the generators
we give the CPU time needed on one core of an Intel(R) Xeon(R) CPU E3-1225 v3
for one update and measurement cycle for $L=32$. The program has been 
compiled with the gcc version 9.3.0 and the \verb+-O2+ optimization.
 We need $0.00198$ s, 
$0.00197$ s, $0.00228$ s, and $0.00307$ s using the SFMT, the 
\verb$ xoshiro256+$, the modified KISS, and the RANLUX generator, respectively.
Note that for one sweep with the local update we need exactly $L^3=32768$ 
random numbers. For the cluster algorithms on average about $77200$ 
random numbers are used in one cycle with $L/4=8$ single cluster and
one wall cluster update. In a simple program that only calls the random 
number generator, the \verb$ xoshiro256+$ and the modified KISS for example 
takes $8 \times 10^{-10}$ and $2 \times 10^{-9}$ 
seconds per call, respectively.  The difference between these numbers
does not fully explain the difference 
in the timings for the whole measurement und update cycle given above. 
This might be explained  by the fact
that the random number generator is inlined in the code, and the 
result of the optimization performed by the compiler depends much on the 
code that is inlined.

\section{Three-dimensional XY model with next to next to nearest neighbor
couplings}
\label{XYnnn}
We performed a preliminary finite size scaling study of the XY model
on the simple cubic lattice 
with next to next to nearest neighbor couplings in addition to the nearest
neighbor one. The reduced Hamiltonian is given by
\begin{equation}
\label{XY}
H = -K_1 \sum_{<xy>}  \vec{s}_x \vec{s}_y - K_3 \sum_{[xy]} \vec{s}_x \vec{s}_y
 \;\; ,
\end{equation}
where $\vec{s}_x$ is a unit vector with two real components. Otherwise, the
notation is the same as in sect. \ref{themodel}. The simulations were actually
performed prior to the study discussed in the main part of this paper. 
Therefore the setup slightly differs from that of the main part of this 
paper. In particular, we varied $K_1$, while keeping $K_3$ fixed. 
For $K_3=0.03$ and $0.05$ we simulated the linear lattices 
$L=8$, $10$, $12$,..., $20$. For $K_3=0.04$, we simulated 
$L=6$, $7$, $8$,..., $20$, $22$, $24$, ..., $30$, $34$, $40$, $50$, ..., $80$.
We performed $3 \times 10^9$ measurement for $L \le 20$. 
In the case of $K_3=0.04$, the number of measurement is decreasing, going 
to larger lattice sizes. For $L =80$, $5.5 \times 10^8$ measurement were
performed. We simulated at good estimates of the critical coupling $K_{1,c}$.
These estimates were iteratively improved with increasing lattice size.
Making use of $(Z_a/Z_p)^*=0.32037(6)$ for the three-dimensional 
XY universality class \cite{myClock}, we get 
$K_{1,c}=0.3931647(10)$, $0.3736005(2)$, and $0.3544282(10)$, 
for $K_3=0.03$, $0.04$, and $0.05$, respectively.

Next, we analyzed $U_4$ at $(Z_a/Z_p)^*=0.32037$ by using the Ansatz
\begin{equation}
\bar{U}_4 = \bar{U}_4^* + \bar{b}(K_3) L^{-\omega} + 
\bar{c}(K_3) L^{-2 + \eta}, 
\end{equation}
using the estimates $U_4^*=1.24296(8)$ and $\omega=0.789(4)$ 
for the three-dimensional XY universality class \cite{myClock}.
For $\omega=0.789$ fixed, we arrive at $\bar{b}=-0.031(2)$, $-0.0045(15)$, 
and $0.021(2)$ for 
$K_3=0.03$, $0.04$, and $0.05$, respectively. The error of $\bar{b}$ 
is dominated by
the uncertainty of $U_4^*$ that we used as input. Linearly interpolating, 
we arrive at $q_3^*=0.113(2)$. In the large $N$-limit, where
$N$ counts the number of components of the spin $\vec{s}_x$, 
$q_{3,N=\infty}^{iso}=q_{3,free}^{iso}=0.125$. Therefore, it is plausible that 
$0.125 < q_{3,XY}^{iso} < q_{3,Ising}^{iso}$, meaning that
$q_3^{iso} > q_3^*$ for the XY model.
Hence, following the argument of sect. \ref{themodel}, we can
not find a model similar to eq.~(\ref{BlumeCapel}),
where both the leading correction to
scaling and the leading violation of isotropy are eliminated.
Still the XY model at $q_3^*$ might be useful, since the spatial anisotropy
should be considerably reduced compared with $q_3=0$.

\end{document}